\documentclass[letterpaper,journal]{IEEEtran}
\usepackage{amsmath,amsfonts}
\usepackage{algorithm}
\usepackage{algpseudocode}
\usepackage{array}
\usepackage[caption=false,font=normalsize,labelfont=sf,textfont=sf]{subfig}
\usepackage{textcomp}
\usepackage{stfloats}
\usepackage{url}
\usepackage{verbatim}
\usepackage{graphicx}
\usepackage{cite}
\usepackage{xcolor}
\usepackage{amssymb}
\usepackage{enumitem}
\usepackage{float}
\usepackage{booktabs, multirow}
\usepackage{tabularx}
\let\originalleft\left
\let\originalright\right
\renewcommand{\left}{\mathopen{}\mathclose\bgroup\originalleft}
\renewcommand{\right}{\aftergroup\egroup\originalright}

\begin{document}

\title{Random Access for LEO Satellite \\ Communication Systems via Deep Learning}

\author{Hyunwoo Lee,~\IEEEmembership{Graduate Student Member,~IEEE,} Ian P.~Roberts,~\IEEEmembership{Member,~IEEE,} \\Jinkyo Jeong,~\IEEEmembership{Member,~IEEE,} and Daesik Hong,~\IEEEmembership{Fellow,~IEEE}
        % <-this % stops a space
\thanks{H. Lee and D. Hong are with the Information Telecommunication Lab (ITL), School of Electrical and Electronic Engineering, Yonsei University, Seoul, South Korea (e-mail: gksdnrh27@yonsei.ac.kr and daesikh@yonsei.ac.kr).

I.~P.~Roberts is with the Wireless Lab, Department of Electrical and Computer Engineering, UCLA, Los Angeles, CA, USA (e-mail: ianroberts@ucla.edu).

J. Jeong is with the Department of Electronic Engineering, Pukyong National University, Busan 48513, South Korea (e-mail: jjk.jeong@pknu.ac.kr).}
}% <-this % stops a space
%\thanks{Manuscript received April 19, 2021; revised August 16, 2021.}}

% \markboth{Journal of \LaTeX\ Class Files,~Vol.~14, No.~8, S~2021}%


\maketitle

\begin{abstract}
% To support massive connectivity in 6G networks, non-terrestrial networks (NTNs) based on low Earth orbit (LEO) satellite constellations have drawn significant attention due to their wide coverage and low-latency communication capabilities. 
Integrating contention-based random access procedures into low Earth orbit (LEO) satellite communication (SatCom) systems poses new challenges, including long propagation delays, large Doppler shifts, and a large number of simultaneous access attempts. 
These factors degrade the efficiency and responsiveness of conventional random access schemes, particularly in scenarios such as satellite-based internet of things and direct-to-device services.
In this paper, we propose a deep learning-based random access framework designed for LEO SatCom systems. 
The framework incorporates an early preamble collision classifier that uses multi-antenna correlation features and a lightweight 1D convolutional neural network to estimate the number of collided users at the earliest stage. 
Based on this estimate, we introduce an opportunistic transmission scheme that balances access probability and resource efficiency to improve success rates and reduce delay. 
Simulation results under 3GPP-compliant LEO settings confirm that the proposed framework achieves higher access success probability, lower delay, better physical uplink shared channel utilization, and reduced computational complexity compared to existing schemes.
\end{abstract}


\section{Introduction}\label{sec_1}
% To achieve truly ubiquitous and seamless connectivity, 6G networks are expected to extend beyond terrestrial infrastructure by incorporating non-terrestrial networks (NTNs). 
% These networks leverage aerial and spaceborne platforms—such as satellites, high-altitude platforms, and unmanned aerial vehicles—to provide broader coverage.
Among the various types of non-terrestrial network platforms, large-scale low Earth orbit (LEO) satellite constellations have emerged as a key enabler, owing to their ability to provide near-global coverage and relatively low latency due to their proximity to the Earth compared to higher-orbit systems~\cite{WCM23_NTNMAG,WCM24_LEOhardware,COMMAG26_TDD}.
% While LEO systems fall under the broader NTN category, they have become a central focus of standardization efforts led by the 3rd Generation Partnership Project (3GPP)~\cite{36.763,38.811,38.821}.
Reflecting this focus, since Release 17, the 3rd Generation Partnership Project (3GPP) has concentrated on adapting terrestrial cellular technologies to tackle the unique challenges of LEO satellite communication (SatCom), such as long communication distances and high mobility~\cite{36.763,38.811,38.821}.

Random access procedures are fundamental components of wireless networks that enable users to initiate connections with base stations (BSs).
These procedures typically involve a sequence of message exchanges to establish uplink synchronization and request transmission resources.
In terrestrial networks such as 4G LTE and 5G NR, contention-based random access is used, where users transmit randomly selected preambles and resolve potential collisions through a multi-step handshake with the BS~\cite{Book_RABO}.
This approach is well-suited for terrestrial environments as it provides a scalable and efficient way to manage sporadic access attempts from a large number of users.
However, adapting these contention-based mechanisms to LEO SatCom systems requires careful consideration of several LEO-specific constraints that significantly impact the performance and reliability of the access procedure.
These constraints are summarized as follows, based on a representative LEO scenario with an altitude of 600\,km and a minimum service elevation angle of 10 degrees~\cite{38.811}:

\begin{itemize}
    \item \textbf{Long propagation delay:} The considerable distance between the satellite and ground users introduces significant propagation delays.  
    Depending on the user’s location within the satellite footprint, the one-way delay ranges from approximately 2\,ms to 6.44\,ms, resulting in a round-trip time (RTT) variation of up to 4.44\,ms~\cite{38.811}.  
    In random access procedures, such delays cause latency in the exchange of signaling messages, which can degrade responsiveness and increase the time required to complete access attempts.  
    Moreover, because the satellite BS (SBS) must receive preambles from all users without prior timing synchronization, it must set a reception window wide enough to account for this RTT uncertainty.
    \item \textbf{Large Doppler shift:} LEO satellites travel at high speeds, approximately 7.56\,km/s, resulting in significant Doppler shifts.  
    For instance, at a 2\,GHz carrier frequency, this can lead to frequency offsets up to 48\,kHz~\cite{38.811}, which is nearly 38 times the 1.25\,kHz subcarrier spacing used in the 5G NR random access procedure~\cite{38.211}.  
    Since the random access waveform is designed under the assumption of small frequency offsets, such a large shift can cause severe inter-carrier interference and significantly degrade preamble detection performance unless compensated at the user side.
    \item \textbf{Large number of simultaneous access attempts:}  
    Due to the large coverage area of LEO satellites compared to terrestrial BSs, the total number of users served simultaneously is expected to be significantly higher.  
    As a result, the number of users attempting random access at the same time can be substantially greater than in terrestrial networks, leading to a higher probability of preamble collisions and access delays.  
    For instance, in a LEO satellite-based internet of things (IoT) scenario, future systems are expected to support up to one billion IoT nodes within a single satellite footprint~\cite{36.763}.  
    In contrast, terrestrial networks, which operate with smaller coverage areas and more densely deployed infrastructure, typically serve between 100{,}000 and 10{,}000{,}000 nodes in total~\cite{TWC21_DLRA, TWC17_NORA, COMST23_LEOMIMO}.  
    This implies that satellite-based systems may need to accommodate 100 to 10{,}000 times more simultaneous access attempts, significantly intensifying contention on the random access channel.
    Notably, even when a phased-array antenna confines contention to a narrow beam, the beam footprint remains much larger than that of a typical terrestrial cell~\cite{COMST23_LEOMIMO}, and thus high contention can still arise within a single beam.
\end{itemize}
To address two of these challenges—namely, large RTT uncertainty and Doppler-induced frequency offset—3GPP Release 17 introduces the use of global navigation satellite system (GNSS) functionality at the user and ephemeris information broadcast by the satellite~\cite{38.811}.
With knowledge of both the user’s position and the satellite's orbit, the user can pre-compensate for the required timing advance (TA) and frequency offset before transmitting its preamble—a known sequence used to initiate random access~\cite{38.811,38.821,36.763}.
This allows reuse of the terrestrial contention-based random access procedure in LEO SatCom systems, in which the user transmits a preamble (Step 1), the BS responds (Step 2), the user sends identity information (Step 3), and the BS acknowledges (Step 4).
However, despite enabling this reuse, the approach suffers from critical limitations in detecting and resolving preamble collisions.
Collisions in Step 1 are not detected at the time of transmission but are only revealed later when the BS fails to decode signals in Step 3.
On the user side, a collision is inferred only when a message is not received within the collision resolution (CR) timer in Step 4.

This mechanism results in delays of at least two RTTs and leads to inefficient use of physical uplink shared channel (PUSCH) resources allocated for Step 3.
Such issues are present in terrestrial networks but exacerbate in LEO systems due to inherently longer delays and a larger number of simultaneous access attempts.
LEO satellite-based IoT scenarios currently serve as a representative example, where the high density of devices leads to frequent access contention.
Similar challenges are also expected in emerging applications such as direct-to-device (D2D) services, where the number of connected handheld devices is projected to grow significantly in the coming years~\cite{AESM25_D2D}.
These observations highlight the need for improved random access mechanisms that can detect collisions early and increase access success rates, particularly in LEO SatCom systems operating under high user density.

The rest of this paper is organized as follows. 
Section~\ref{sec_2} provides an overview of related work aimed at resolving preamble collisions and outlines the main contributions of this paper.
Section~\ref{sec_3} describes the conventional random access procedure for LEO SatCom systems. 
The proposed random access framework is presented in Section~\ref{sec_4}. Section~\ref{sec_5} provides simulation results to demonstrate the effectiveness of the proposed framework and discusses key observations.
Finally, Section~\ref{sec_6} concludes the paper.

\section{Related Work and Contributions}\label{sec_2}
\subsection{Related Work}
Contention-based random access serves as a fundamental mechanism in cellular systems.
A central challenge in this framework is resolving preamble collisions, as they directly impact the success probability and latency of the access procedure.
To address this challenge, extensive research has been conducted, with prior efforts broadly categorized into two directions:
(1) reducing the likelihood of preamble collisions and
(2) resolving preamble collisions once they occur.

\subsubsection{\textbf{Reducing the likelihood of preamble collisions}}
To alleviate preamble collisions in contention-based access, various techniques have been proposed to reduce the probability that multiple users select the same preamble simultaneously.  
Key approaches explored in the literature include access class barring (ACB)~\cite{TVT18_ACB1, TVT23_ACB2, TWC24_ACB3}, adaptive random backoff strategies~\cite{TVT16_BO1, TN18_BO2, TMC24_BO3}, and non-orthogonal preambles~\cite{TVT18_NP1, TVT21_NP2, TWC21_NP3, CL17_NP4, TVT23_TwoStage}.

In the ACB scheme, the number of access attempts per random access channel (RACH) slot is regulated to avoid excessive collisions.  
Before transmitting a preamble, each user performs an ACB check and proceeds only if it passes; otherwise, the user defers the attempt to the next available RACH slot~\cite{TVT18_ACB1,TVT23_ACB2,TWC24_ACB3}.  
% In~\cite{TVT18_ACB1}, the authors analyzed the performance of ACB under 3GPP-specified random access scenarios and proposed appropriate configuration parameters.  
% Further studies explored the use of machine learning~\cite{TVT23_ACB2} and reinforcement learning~\cite{TWC24_ACB3} to optimize ACB parameters.
Similarly, many works have focused on adjusting the backoff behavior of users after collisions.  
In~\cite{TVT16_BO1, TN18_BO2, TMC24_BO3}, dynamic backoff schemes were proposed that adapt the uniform backoff window based on observed collision levels.  
% In~\cite{TMC24_BO3}, the authors showed that such dynamic backoff control helps alleviate congestion and reduce the age of information.
While ACB and backoff adjustment help mitigate RACH congestion, they require tracking the backlog state of all users.  
These approaches are less attractive in LEO SatCom systems, where the service region changes dynamically due to satellite movement.

One of the most effective ways to reduce preamble collisions is to increase the number of available preambles~\cite{TVT18_NP1, TVT21_NP2, TWC21_NP3, CL17_NP4, TVT23_TwoStage}.
% In LTE and NR systems, preambles are typically generated from Zadoff–Chu (ZC) sequences with different root indices~\cite{Book_RABO}.
% However, the number of orthogonal preambles is fundamentally limited by the sequence length and the correlation properties of ZC sequences.
The authors of~\cite{TVT18_NP1, TVT21_NP2} proposed a concatenated preamble structure that forms a composite sequence by combining two Zadoff-Chu (ZC) sequences generated from different roots.
This combinatorial design increases the number of unique preambles without requiring longer sequences or additional time-frequency resources.
Although this expanded preamble space reduces collision probability, it also leads to increased non-orthogonality, resulting in higher interference as the user density grows.
To mitigate such interference, successive interference cancellation techniques have been employed in~\cite{TWC21_NP3, CL17_NP4, TVT23_TwoStage}.
While these methods help detect non-orthogonal preambles more effectively, they rely on iterative cancellation procedures, which entail high computational complexity.

Recent learning-based studies have investigated access-control policies for LEO satellite networks to mitigate contention under non-stationary topology and time-varying multi-satellite coverage. 
For example, \cite{TWC23_RL1} proposes an emergent contention-based random access protocol (eRACH) learned via multi-agent DRL, where distributed agents adapt their access behaviors to evolving network states. 
Similarly, \cite{IoTJ25_RL2} formulates user access as a long-term decision problem under time-varying coverage and develops a decentralized multi-agent RL mechanism to maximize long-term throughput. 
In addition, some DRL studies in the satellite context focus on higher-layer network control such as transport-layer congestion control in space--air--ground integrated networks (SAGINs), rather than contention-based random access \cite{TNSE26_RL3}. 
Overall, these RL-based approaches aim to reduce congestion through long-term access adaptation, thereby lowering the likelihood of persistent contention in random access procedures.

\subsubsection{\textbf{Resolving preamble collisions once they occur}}
When multiple users select the same preamble and transmit it simultaneously, the BS cannot distinguish between them and responds with the same message.
As a result, all those users receive the same grant for Step 3, causing a collision that leads to random access failure.
Several approaches have been proposed to handle such collisions after they occur, aiming to recover from contention rather than prevent it in advance.
These include transmission control~\cite{PIMRC17_BCCR, IoTJ20_MSG3B}, non-orthogonal random access~\cite{TCOM23_NOMA1, IoTJ23_NOMA2, TMC24_NOMA3}, and early preamble collision detection schemes~\cite{TWC17_NORA, TVT18_eCD2, TVT21_SeCD, TVT25_MTeCD, TWC21_DLRA, TWC23_DLDouble}.

A user-side approach in~\cite{PIMRC17_BCCR} enables early collision detection by having users broadcast short contention signals before Step 3 transmission.
While effective in terrestrial settings, this method is unsuitable for LEO SatCom due to wide coverage and limited sensing among users.
Alternatively, a BS-centric scheme called Step 3 barring was proposed in~\cite{IoTJ20_MSG3B}, where the BS estimates collision levels and broadcasts a uniform barring probability to all users.
Although compatible with LEO SatCom, this approach cannot differentiate collided from non-collided users, causing unnecessary barring and increased access delay.

In~\cite{TCOM23_NOMA1, IoTJ23_NOMA2, TMC24_NOMA3}, non-orthogonal random access schemes were proposed where users sharing the same Step 3 grant transmit simultaneously using power-domain NOMA.
This enables multiple users to share uplink resources through superimposed transmissions and works well in terrestrial networks with diverse channel gains.
In LEO SatCom, however, the long transmission distance yields minimal power disparities, making such separation challenging and limiting their effectiveness~\cite{ICASSP23_NOMA}.

Several studies have sought to detect preamble collisions at the BS immediately after receiving Step 1 transmissions.
A common approach is to analyze the correlation output to identify multiple time of arrival (ToA) peaks~\cite{TWC17_NORA, TVT18_eCD2}, which works in terrestrial networks but becomes ineffective in LEO SatCom due to TA pre-compensation using GNSS and ephemeris information~\cite{36.763}.
To overcome this limitation, alternative schemes that do not rely on ToA differences have been proposed~\cite{TVT21_SeCD, TVT25_MTeCD, TWC21_DLRA, TWC23_DLDouble}.
For example, a prime-based cyclic shift scheme~\cite{TVT21_SeCD} enables fast collision detection but is sensitive to RTT uncertainty and fading.
A multi-threshold method~\cite{TVT25_MTeCD} instead uses empirically derived thresholds, but lacks robustness under varying channel conditions.

There have also been attempts to apply deep learning for early preamble collision detection~\cite{TWC21_DLRA, TWC23_DLDouble}.
In~\cite{TWC21_DLRA}, a neural network was used to estimate the number of colliding users and extract TA values, enabling the BS to broadcast multiple Step 3 grants.
However, in LEO SatCom, the 3GPP-mandated GNSS-based TA pre-compensation aligns the arrival timing of preambles at the satellite, making the per-user identification on which this approach relies infeasible.
While the method achieves strong detection performance in terrestrial settings, its fully connected network architecture incurs high computational complexity.
In~\cite{TWC23_DLDouble}, a double contention scheme was proposed in which a neural network allocates two Step 3 grants when two users collide, but it also suffers from high complexity, limited scalability beyond two users, and the need for dynamically adjustable Step 3 resources, which conflicts with current standards~\cite{Book_RABO}.
In contrast, our proposed framework resolves contention probabilistically based on classifier outputs while explicitly accounting for classification uncertainty, without modifying baseline 3GPP resource allocation.
Table~\ref{tab:dl_comparison} provides a side-by-side summary of these differences along four key dimensions.

\begin{table}[!t]
\caption{Comparison between the proposed framework and prior deep-learning-based random access schemes.}
\label{tab:dl_comparison}
\centering
\renewcommand{\arraystretch}{1.15}
\footnotesize
\begin{tabular}{|p{2.0cm}|p{1.5cm}|p{1.5cm}|p{1.6cm}|}
\hline
 & \textbf{DRA~\cite{TWC21_DLRA}} & \textbf{DCRA~\cite{TWC23_DLDouble}} & \textbf{Proposed} \\
\hline
Contention resolution & Deterministic (TA-based) & Deterministic (multiple grants) & Stochastic (analytically optimized) \\
\hline
Classifier uncertainty modeling & No & No & Yes \\
\hline
Additional PUSCH per collision & Yes & Yes & No \\
\hline
3GPP NR/NTN compatibility & Infeasible (TA-based identification fails under GNSS pre-compensation) & Requires non-standard PUSCH allocation & Backward-compatible \\
\hline
\end{tabular}
\end{table}

\subsection{Contributions}
Most existing studies that address the preamble collision problem still suffer from several limitations.
These include unnecessary delay, high computational complexity at the BS, increased complexity at the user, reduced PUSCH resource efficiency, and limited applicability in LEO SatCom systems.

In this paper, we propose a deep learning-based random access framework that overcomes these limitations and is well-suited for LEO SatCom systems.
Deep learning is employed to enable early detection of preamble collisions and to estimate the number of colliding users, a task that cannot be performed effectively without learning-based techniques.
This capability is leveraged to enhance collision resolution and improve random access performance.
The main contributions of this work are summarized as follows.
\begin{itemize}
    \item We propose a deep learning-based early preamble collision classifier tailored for LEO SatCom systems, which is not a direct adaptation of existing terrestrial classifiers but reflects NTN-specific design considerations.
    The classifier leverages per-antenna correlation values (rather than averaged values) as input features, enabling early estimation of the collision multiplicity (i.e., the number of users selecting the same preamble) during the initial stage of the random access process.
    To satisfy practical satellite payload constraints, a lightweight 1D convolutional neural network architecture with a kernel size matched to the NTN channel delay spread is adopted to reduce on-board computation and memory requirements while maintaining high classification accuracy.
    This lightweight design enables fast contention handling and supports practical NTN deployment under stringent power and latency budgets.
    
    \item We propose a collision-aware opportunistic Step 3 transmission mechanism to improve random access performance.
    The scheme exploits the estimated collision multiplicity from the proposed classifier and derives, in closed form, a preamble-index-wise transmission probability that maximizes the random access success probability while explicitly accounting for the classifier's estimation uncertainty.
    Unlike prior deep learning-based schemes that adopt deterministic resource allocation upon collision detection, the proposed mechanism resolves contention probabilistically without allocating additional uplink resources.
    By enabling probabilistic Step 3 attempts rather than a strict backoff, the proposed mechanism improves the random access success opportunity and mitigates delay and uplink resource waste caused by unnecessary transmissions.

    \item We provide an NTN-compliant, system-level validation under a 3GPP-aligned LEO SatCom scenario and extract actionable design insights.
    Rather than merely presenting simulation plots, the evaluation substantiates the feasibility of integrating the proposed framework into a 3GPP-aligned random access procedure and quantifies its system-level gains over benchmark schemes.
    In addition, the results identify operating regimes where the proposed framework yields clear benefits and offer practical guidelines for key system parameters such as active-user load and received SNR regimes.
\end{itemize}

\section{Preliminaries}\label{sec_3}
%\subsection{LEO satellite-based NTN IoT systems}
%We consider the LEO SET-2 configuration for NTN IoT systems~\cite{36.763}. 
%In this configuration, LEO satellites orbit at an altitude of 600 km with a velocity of 7.56 km/s. 
%The minimum elevation angle for service is 10 degrees, allowing the satellite to serve nodes located within the 10 to 90-degree elevation range.
%Additionally, we assume that IoT nodes are equipped with GNSS functionality. 
%Thanks to GNSS functionality and satellite ephemeris information, TA and frequency offset can be pre-compensated, which effectively mitigates the impact of large RTT uncertainty and significant Doppler shift. 
%This assumption is in line with the enhancements introduced in 3GPP Release 17, where the accuracy of TA and frequency offset pre-compensation is considered highly reliable~\cite{38.133,38.821}. 
%As a result, the contention-based random access framework used in terrestrial cellular systems can be employed to LEO satellite-based NTN systems without significant modifications. 

\begin{figure}[t]
    \centering
    \includegraphics[width=1\columnwidth]{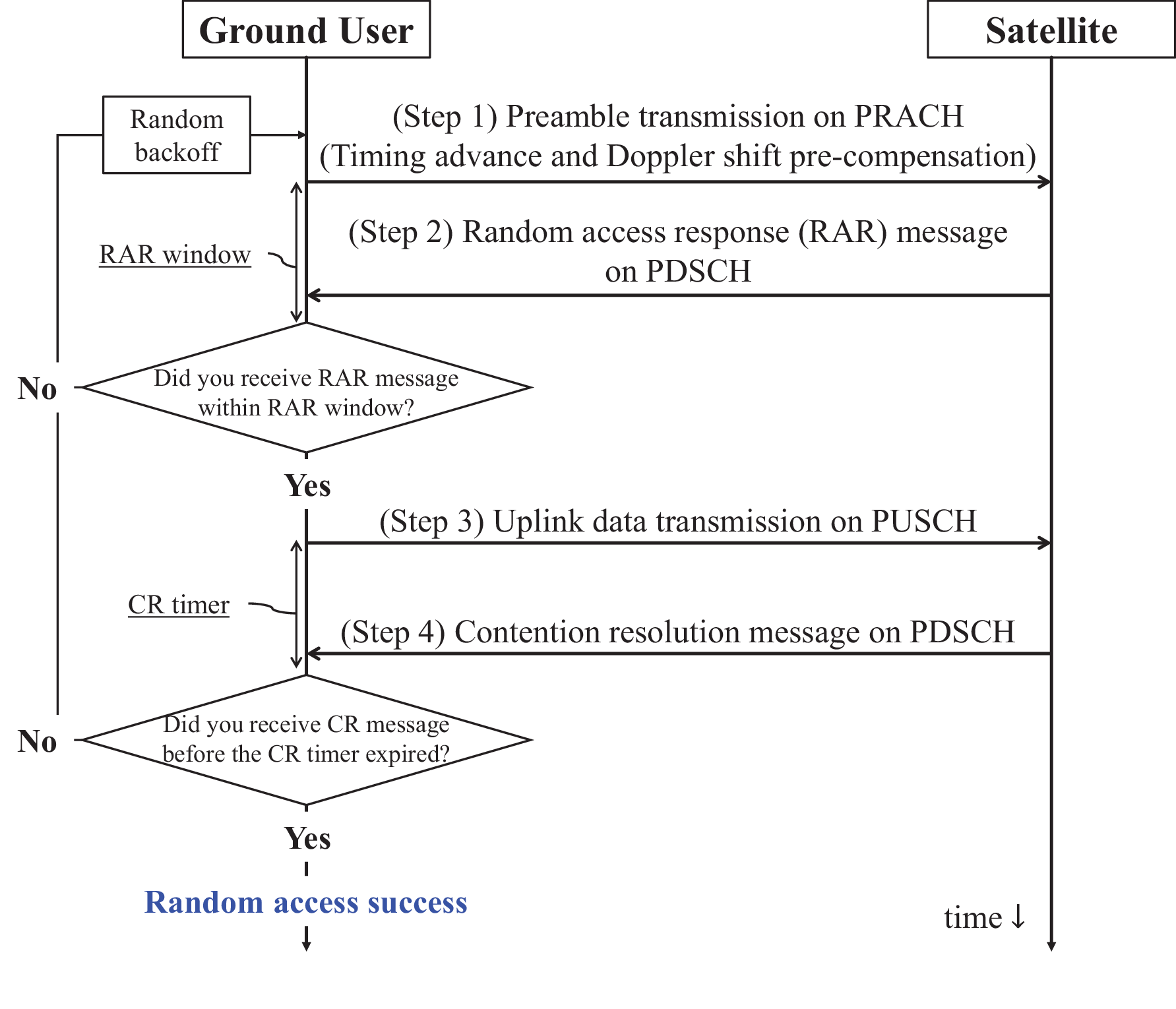}
    \caption{Conventional contention-based random access procedure for LEO SatCom systems.}
    \label{fig:conv_framework}
\end{figure}

%\subsection{Random access procedure for NTN IoT systems}
This work builds upon the contention-based random access framework defined by 3GPP for SatCom systems~\cite{38.811}.
Before introducing our proposed framework, it is therefore useful to outline the 3GPP-specified procedure to establish notation and terminology.
As illustrated in Fig.~\ref{fig:conv_framework}, the contention-based random access framework for LEO SatCom systems consists of the following four steps.

\textbf{(Step 1) Preamble Transmission:} Since 4G LTE systems, a ZC sequence has been used as the random access preamble in 3GPP~\cite{Book_RABO}, and is expressed as
\begin{equation}
    z_r[n]=\exp{\left[\frac{-j\pi rn\left(n+1\right)}{N_{\mathrm{ZC}}}\right]}, \quad n=0,1,...,N_{\mathrm{ZC}}-1,
    \label{eq:ZC}
\end{equation}
where $r\in\left\{1,...,N_\mathrm{R}\right\}$ represents the root index, $N_\mathrm{R}$ and $N_{\mathrm{ZC}}$ denote the number of roots and the sequence length respectively. 
One of the key advantages of the ZC sequence lies in its \textit{cyclic cross-correlation property}, which ensures that sequences with different root indices exhibit a constant cross-correlation as
\begin{equation}
    \left|c_{rs}\right[m]|=\left|\frac{1}{\sqrt{N_{\mathrm{ZC}}}}\sum^{N_{\mathrm{ZC}}-1}_{n=0}{z_r[n]z_s^*\left[\left(n+m\right)\bmod N_{\mathrm{ZC}}\right]}\right|=1.
\end{equation}
Here, $c_{rs}[m]$ denotes the \emph{normalized} cyclic cross-correlation, where the unnormalized correlation $\sum_{n=0}^{N_{\mathrm{ZC}}-1} z_r[n] z_s^*[(n+m)\bmod N_{\mathrm{ZC}}]$ has magnitude $\sqrt{N_{\mathrm{ZC}}}$.
The ZC sequence also has the following \textit{cyclic auto-correlation property} in which the auto-correlation function takes the form of the Dirac delta function $\delta[\cdot]$ as
\begin{equation}
    \begin{aligned}
        \left|c_{rr}\right[m]| & =\left|\frac{1}{\sqrt{N_{\mathrm{ZC}}}}\sum^{N_{ZC}-1}_{n=0}{z_r[n]z_r^*\left[\left(n+m\right)\bmod N_{\mathrm{ZC}}\right]}\right| \\ 
        & =\sqrt{N_{\mathrm{ZC}}}\cdot\delta[m] = 
        \begin{cases}
        \sqrt{N_{\mathrm{ZC}}}, & m = 0, \\
        0, & m \neq 0.
        \end{cases}
    \end{aligned}
\end{equation}
These correlation properties allow each preamble to be uniquely identified by its root index $r$ and cyclic shift index $i$.
To enable multiple users to transmit distinct preambles derived from the same root sequence without causing mutual interference, a cyclic shift size $N_{\mathrm{CS}}$ is introduced.
This parameter ensures that the correlation peaks corresponding to different cyclically shifted sequences do not overlap within the correlation window at the receiver.
Accordingly, a preamble sequence is defined as
\begin{equation}
    z_{r,i}\left[n\right]=z_r[\left(n+iN_{\mathrm{CS}}\right)\bmod N_{\mathrm{ZC}}], \quad n=0,1,...,N_{\mathrm{ZC}}-1,
\end{equation}
where $N_{\mathrm{CS}}$ represents the cyclic shift size and determines the range of $i\in\left\{0,...,\left(\lfloor N_{\mathrm{ZC}}/N_{\mathrm{CS}}\rfloor-1\right)\right\}$. 
Since there are $N_\mathrm{R}$ root indices, the total number of available preambles is $N_{\mathrm{PA}}=N_\mathrm{R}\cdot \lfloor N_{\mathrm{ZC}}/N_{\mathrm{CS}}\rfloor$. 
Users randomly select one of these available preambles and transmit it to the SBS when requesting random access. 
Given the large RTT uncertainty and Doppler shift inherent in satellite links, users are assumed to apply pre-compensation for TA and frequency offset prior to transmission~\cite{38.811}.
As investigated in 3GPP, such compensation can be accomplished by GNSS-based synchronization, which can achieve high estimation accuracy~\cite{38.133,3GPP_FreqEst}.

\begin{figure}[t]
    \centering
    \includegraphics[width=1\columnwidth]{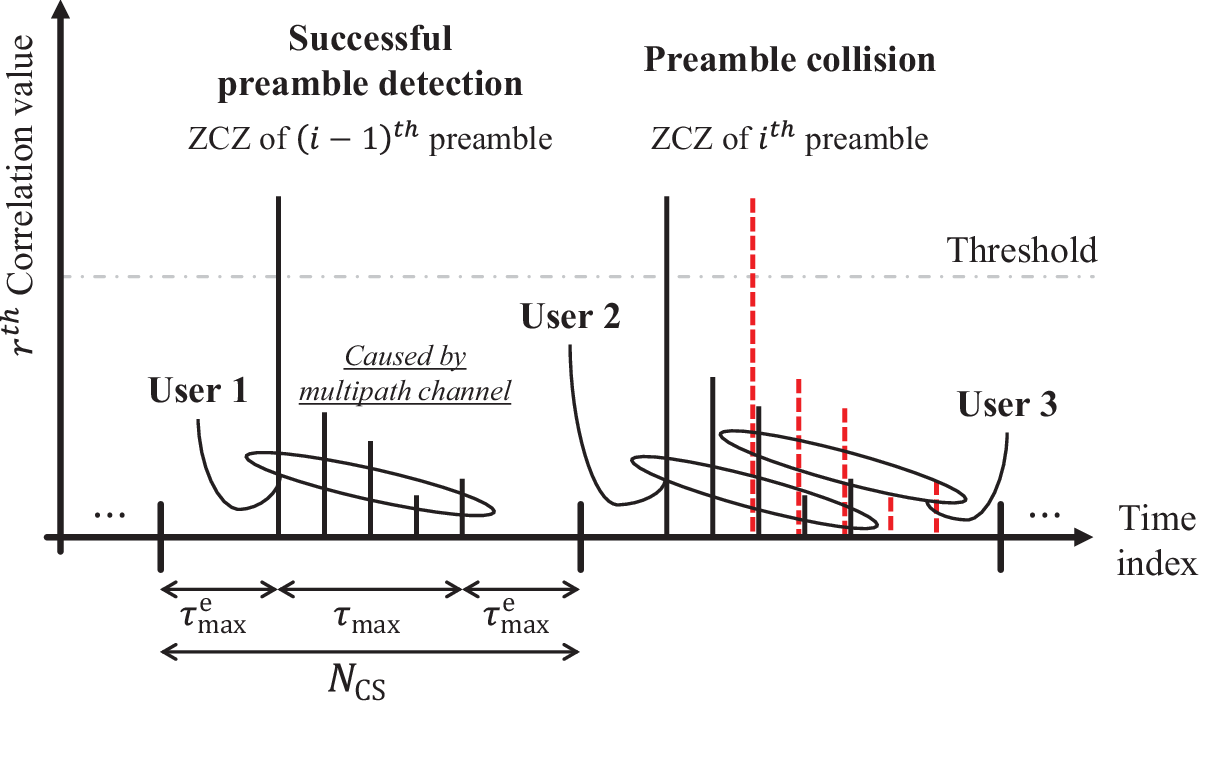}
    \caption{Preamble detection process in the SBS. The SBS computes the correlation values for each predefined ZCZ and then detects the presence of preambles based on a threshold.}
    \label{fig:preamble_detection}
\end{figure}

\textbf{(Step 2) Preamble Detection and Random Access Response Transmission:} After users transmit their preambles, the SBS receives the composite signal and computes the correlation values between the received signal and each of the $N_{\mathrm{R}}$ root ZC sequences to identify the transmitted preambles, as illustrated in Fig.~\ref{fig:preamble_detection}.
Each root sequence is associated with multiple cyclically shifted versions, and each of these shifted sequences defines a preamble.
To enable accurate detection, a zero correlation zone (ZCZ) is configured for each root sequence, within which the autocorrelation of the ZC sequence is ideally zero, minimizing mutual interference among different preambles.
The correlation value is computed within this ZCZ region, and if the maximum value exceeds a predefined threshold, the corresponding preamble is considered to be detected.
In this work, the ZCZ length is set equal to the cyclic shift size $N_{\mathrm{CS}}$, ensuring that each user’s preamble falls within a distinct non-overlapping zone and can be reliably detected by the SBS.
In terrestrial networks, the ZCZ length is typically determined based on the maximum expected timing misalignment, which includes the maximum TA and the channel delay spread.
However, in LEO SatCom systems, users perform TA pre-compensation before transmission, which alters the delay characteristics.
As a result, the design of ZCZ in SatCom must consider this pre-compensation behavior, and the required ZCZ length is derived accordingly.
\begin{equation}
    N_{\mathrm{CS}}\geq\left(\tau_{\mathrm{max}}+2\tau_{\mathrm{max}}^\mathrm{e}\right),
    \label{eq:ZCZ_condition}
\end{equation}
where $\tau_{\mathrm{max}}$ represents the maximum channel delay spread, and $\tau_{\mathrm{max}}^\mathrm{e}$ denotes the worst-case TA pre-compensation error. 
After detecting the preamble, the SBS transmits a RAR message, which includes the detected preamble index, a TA command to compensate for TA error, and a PUSCH resource grant for Step 3. 
Since the SBS detects preambles solely based on the correlation values of the received signals, it can detect the presence of a preamble but cannot determine how many users transmitted that particular preamble.

\textbf{(Step 3) Uplink Data Transmission:} If users successfully receive the RAR message from the SBS within the predefined RAR window, they transmit the radio resource control (RRC) connection request message using the PUSCH resource grant included in the received RAR message. 
If multiple users have transmitted the same preamble and received the same PUSCH resource grant, they will inadvertently transmit their RRC connection request messages over the same PUSCH resource.

\textbf{(Step 4) Contention Resolution Message Transmission:} If the SBS successfully decodes the RRC connection request message, it transmits a CR message containing the user identifier (ID). 
If multiple users simultaneously use the same PUSCH resource, the SBS would presumably fail to decode the RRC connection request messages in that resource and would thus not respond to those requests. 
A user that receives the CR message within the predefined CR timer sends back a positive acknowledgement (ACK). 
In contrast, users that do not receive the CR message before the CR timer expires reinitiate the random access procedure after some delay based on a predefined random backoff policy~\cite{Book_RABO}.

Based on this conventional scheme, a preamble collision that occurs during Step 1 can only be detected by the user after Step 4. 
In other words, the user must wait for a duration equivalent to two times the RTT before reattempting access after a failed random access attempt. 
Furthermore, in the event of a preamble collision, the same PUSCH resources are assigned to multiple users, causing a failure in decoding the Step 3 message at the BS and leading to significant PUSCH resource waste.
Therefore, if preamble collisions can be detected at a step earlier than Step 4, both the random access delay and PUSCH resource inefficiency can be effectively mitigated.
We address this in the scheme proposed next.

\section{Proposed Random Access Framework}\label{sec_4}
\begin{figure}[t]
    \centering
    \includegraphics[width=1\columnwidth]{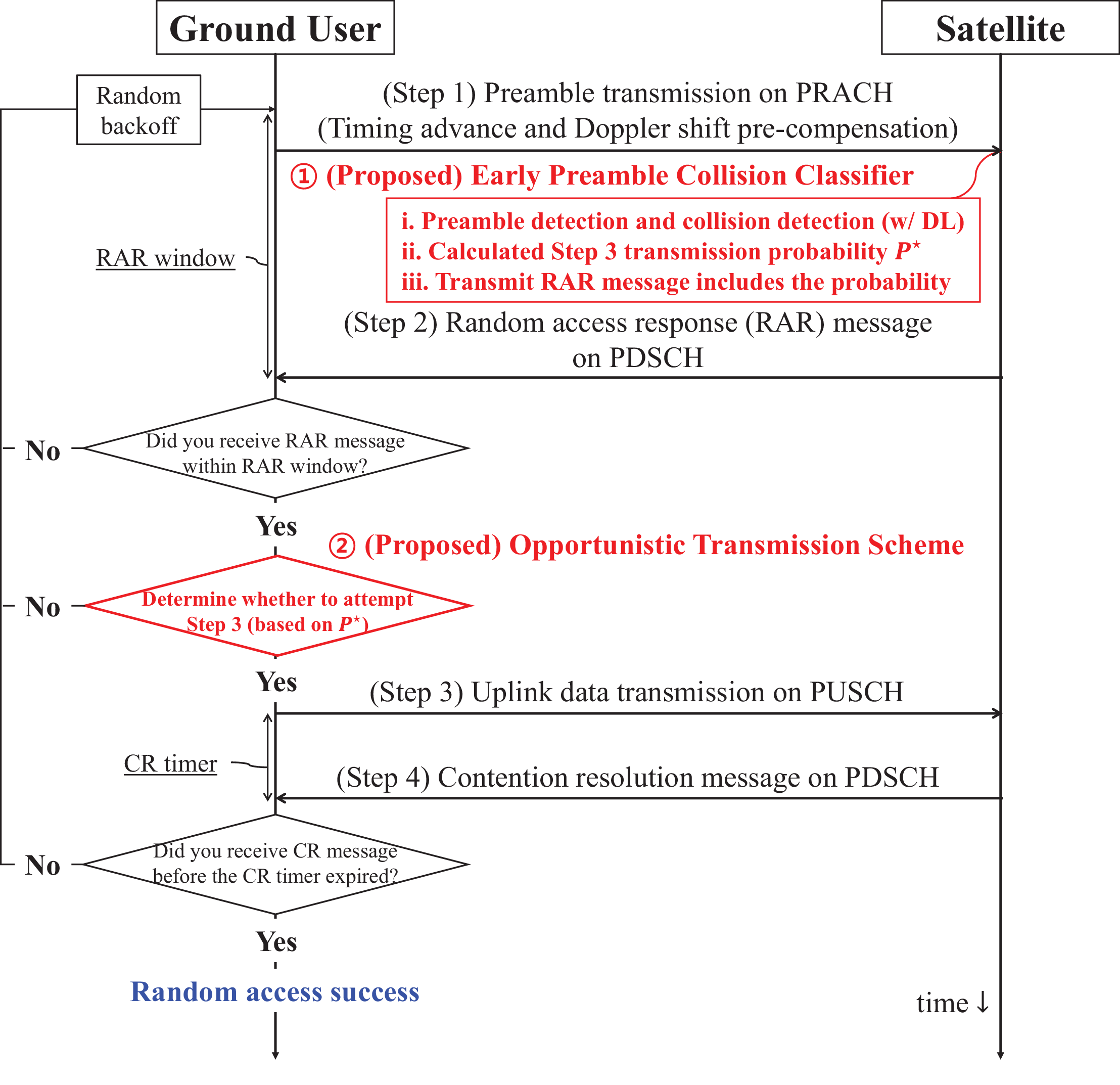}
    \caption{Proposed random access framework for LEO SatCom systems. The framework modifies the conventional random access procedure by incorporating an early preamble collision classifier and an opportunistic transmission scheme for Step 3.}
    \label{fig:prop_framework}
\end{figure}

Building on the observation that preamble collisions in LEO random access cause excessive delay and PUSCH resource waste, we propose an enhanced framework specifically designed to address these issues.
As illustrated in Fig.~\ref{fig:prop_framework}, the proposed framework modifies the legacy 4-step procedure by incorporating two key features:
(1) a deep learning-based preamble collision classifier that enables early detection before Step 4, and
(2) an opportunistic transmission scheme for Step 3 that reduces unnecessary PUSCH usage in the presence of collisions.
These enhancements aim to significantly reduce random access delay and improve PUSCH utilization efficiency in LEO environments.
\subsection{Preamble Transmission and Reception}
In the proposed framework, each user randomly selects a preamble from the available set and applies TA and Doppler shift pre-compensation prior to transmission.
The transmit signal of the $d$th user is thus expressed as
\begin{equation}
    s_d\left[n\right]=\sqrt{P_d}z_{r_d}\left[\left(n+i_dN_{\mathrm{CS}}+\hat{\tau}_d\right)\bmod N_{\mathrm{ZC}}\right]e^{-j2\pi \hat{f}_dnT_\mathrm{s}},
\end{equation}
where $P_d$ is the transmit power, and $\hat{\tau}_d$ and $\hat{f}_d$ denote the estimated TA and Doppler shift, respectively.
$T_\mathrm{s}$ denotes the sampling period. 
We consider a multi-path channel model consisting of $L_d$ paths for the $d$th user.
The SBS is equipped with $N_\mathrm{ant}$ antennas, each of which receives the superposition of signals reflected through multiple paths.
Each path is characterized by a channel coefficient $h_{d,\ell}^j$ and a propagation delay $\tau_{d,\ell}^j$ at the $j$th antenna.
Over the short random-access time scale, we assume a quasi-static multipath channel. 
This is because the considered satellite channel is line-of-sight (LoS)-dominant with negligible scattering near the satellite, and the device is assumed to be stationary during an access attempt~\cite{38.811}.
The signal received at antenna $j$ of the SBS is modeled as
\begin{equation}
    y_j\left[n\right]=\sum^D_{d=1}\sum^{L_d}_{\ell=1}h_{d,\ell}^js_d\left[n+\tau_{d,\ell}^j+\tau^\mathrm{e}_d\right]e^{-j2\pi f_d^\mathrm{e}nT_\mathrm{s}}+w\left[n\right],
\end{equation}
where $D$ denotes the total number of active users that attempt random access in the considered random access occasion, $\tau_d^\mathrm{e}$ and $f_d^\mathrm{e}$ represent the residual errors in timing and frequency pre-compensation for user $d$, and $w[n]$ is additive white Gaussian noise with zero mean and variance $\sigma_w^2$.
Following the conventional detection procedure, the SBS computes the correlation between the received signal and each ZC sequence. 
The correlation at lag $m$ for root index $r$ at antenna $j$ is given by
\begin{equation}
    \left|c_r^j\left[m\right]\right|=\left|\frac{1}{N_{\mathrm{ZC}}}\sum^{N_{\mathrm{ZC}}-1}_{n=0}y_j[n]z_r^*\left[\left(n+m\right)\bmod N_{\mathrm{ZC}}\right]\right|.
    \label{eq:correlation}
\end{equation}

\subsection{Preamble Collision Classifier}
As mentioned, collisions are not detected by traditional schemes until Step 3.
To detect preamble collisions at an early stage, we design a deep learning-based classifier that analyzes the correlation values in~\eqref{eq:correlation}.
Here, $K$ is a design parameter that specifies the maximum collision multiplicity explicitly distinguished by the classifier, and it is independent of the total number of active users $D$ in the system.
For a given root index and cyclic shift, the classifier considers a total $(K+1)$ classes defined as follows:
\begin{itemize}
    \item \textbf{Class 0:} Idle case, where no user has used the preamble.
    \item \textbf{Class 1:} Single user case, where exactly one user has used the preamble (no collision).
    \item \textbf{Class $\boldsymbol{k}$ ($\boldsymbol{2 \le k \le K-1}$):} Exactly $k$ users have used the preamble (collision).
    \item \textbf{Class $\boldsymbol{K}$:} $K$ or more users have used the preamble (collision).
\end{itemize}
The advantage of post-processing the correlation output, as opposed to relying solely on traditional threshold-based detection, is that it not only identifies whether a collision has occurred but also estimates the collision multiplicity up to the designed cut-off $K$ (with class $K$ representing $K$ or more users).

\begin{figure}[!t]
    \centering
    \includegraphics[width=1\columnwidth]{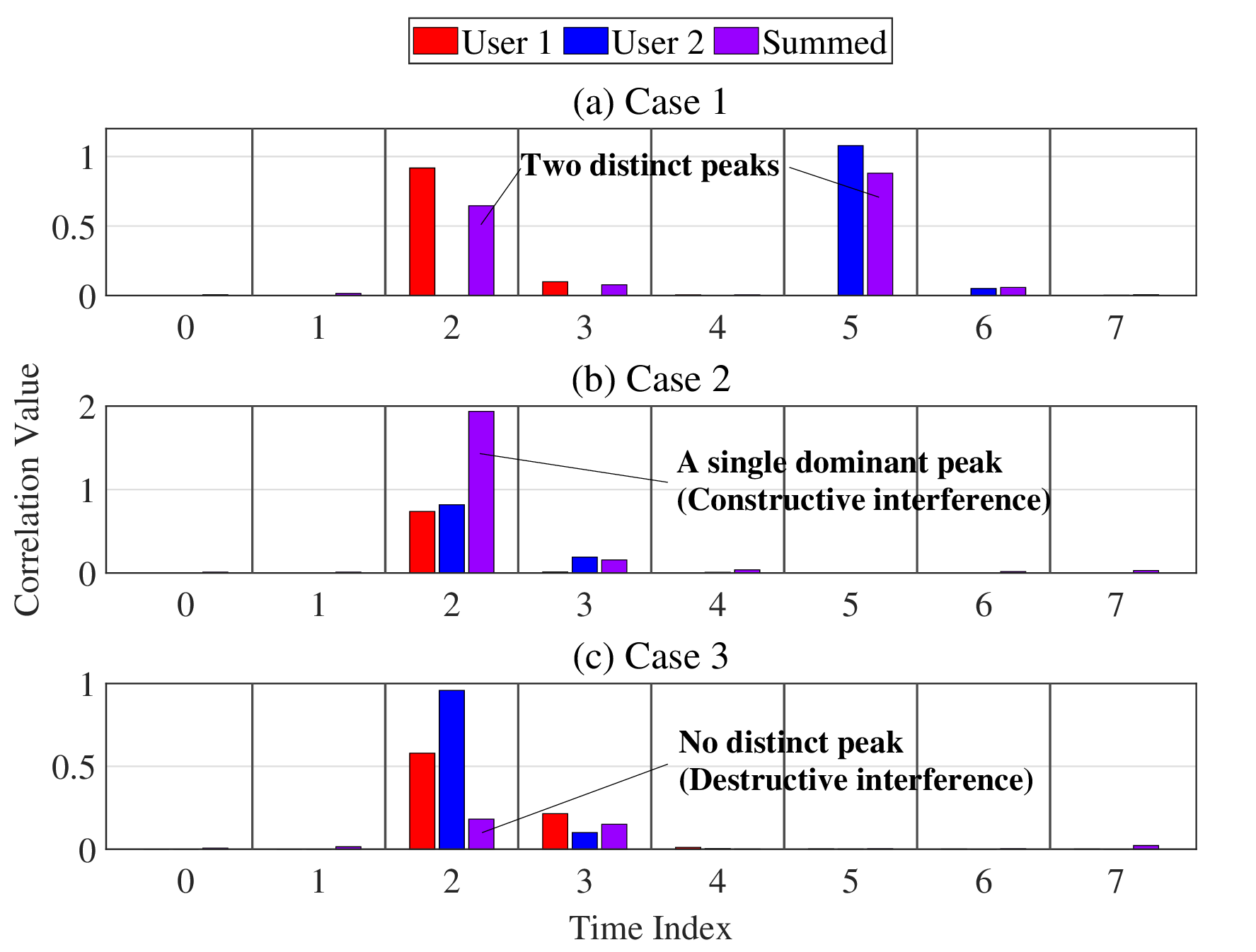}
    \caption{Representative correlation patterns within one ZCZ for a two-user preamble collision: (a) Case~1: two separated peaks (distinct arrivals), (b) Case~2: a single dominant peak due to constructive superposition, and (c) Case~3: no prominent peak due to destructive superposition.}
    \label{fig:ZCZ_cases}
\end{figure}

In realizing such a classifier, our goal is to propose a framework that can be easily integrated into legacy random access procedures by maintaining the core signaling of the conventional random access framework.
Consequently, the input data for the classifier should be correlation values~\eqref{eq:correlation}. 
Since each user independently selects a preamble and collisions across different preambles are independent events, we use a single ZCZ as the input data.
A key point to note, as previously mentioned, is that in SatCom scenarios, TA is pre-compensated by the users during the preamble transmission stage. 
As shown in~\eqref{eq:ZCZ_condition}, the ZCZ size in SatCom scenarios is expected to be considerably smaller than in terrestrial networks, primarily because the otherwise large RTT uncertainty is mitigated through TA pre-compensation.
For example, while the typical ZCZ size in terrestrial networks is approximately 24 samples~\cite{TWC21_DLRA}, it is assumed to be only 8 samples in SatCom scenarios~\cite{TVT23_TwoStage,TVT25_MTeCD}.
This limited ZCZ size imposes a fundamental constraint on the amount of information that can be extracted for preamble collision detection.
More specifically, the smaller number of samples reduces the temporal resolution available to distinguish multiple colliding signals.
As a result, subtle differences in time alignment and energy distribution between users become harder to capture, making it more challenging to detect collisions based solely on the correlation values.
In particular, even for the same collision multiplicity, the resulting correlation sequence within a short ZCZ can vary significantly depending on the relative arrival timing and phase among users, which may create ambiguous observations for simple rule-based detection.

To make this point intuitive, Fig.~\ref{fig:ZCZ_cases} illustrates representative correlation patterns within one ZCZ for a two-user collision, where two users select the same preamble.
In Case~1 (Fig.~\ref{fig:ZCZ_cases}(a)), two sufficiently separated arrivals produce two distinguishable peaks, making the collision visually apparent.
In Case~2 (Fig.~\ref{fig:ZCZ_cases}(b)), closely aligned arrivals can constructively combine into a single dominant peak, which may resemble a collision-free single-user case.
In Case~3 (Fig.~\ref{fig:ZCZ_cases}(c)), destructive superposition suppresses prominent peaks, which can even resemble an idle case.
These examples show that collision-induced signatures are not necessarily captured by a single peak-based statistic and that robust collision inference requires exploiting \emph{local} structures in the ZCZ.

This challenge motivates leveraging additional observations beyond a single correlation trace.
Fortunately, LEO satellites employ multi-antenna systems~\cite{COMST23_LEOMIMO}, which provide multiple observations of the transmitted preambles.
In terrestrial networks, the correlation values obtained from multiple antennas or ports are typically averaged for threshold-based detection~\cite{Book_RABO}.
Similarly, existing deep learning-based schemes have also used the averaged correlation values as input data~\cite{TWC21_DLRA,TWC23_DLDouble}.
Averaging enables receive antenna diversity but at the cost of reduced input dimensionality, which hinders full utilization of the original signal information.
To fully exploit the information contained within preamble signals received across multiple antennas, we use the correlation values from each antenna as input data for the classifier.
This preserves the spatial and temporal diversity in the received signals, providing the classifier with richer information for more accurate collision detection.

\begin{figure*}[t]
    \centering
    \includegraphics[width=2\columnwidth]{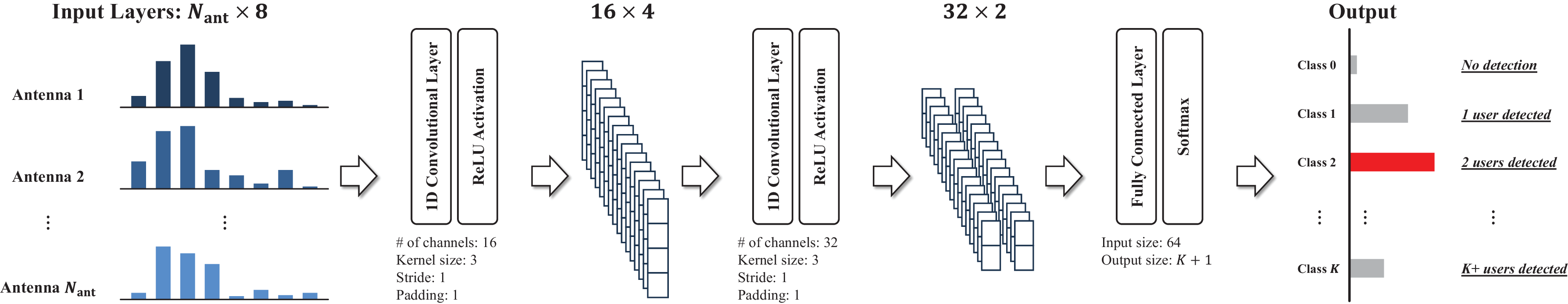}
    \caption{Structure of the proposed early preamble collision classifier. The classifier uses antenna-wise correlation values as input and consists of two 1D convolutional layers followed by a fully connected layer.}
    \label{fig:classifier}
\end{figure*}

However, using antenna-wise correlation values increases the input dimensionality, which requires an appropriate feature extractor to avoid overfitting and to robustly capture collision-induced signatures.
More importantly, the discussion in Fig.~\ref{fig:ZCZ_cases} indicates that collisions manifest as \emph{localized} and sometimes ambiguous structures within a short ZCZ, where the informative cues appear around a few adjacent lags rather than across the entire correlation sequence.
Therefore, we employ 1D convolutional layers to explicitly learn such local patterns and their spatial consistency across antennas, while providing robustness to small timing shifts within the ZCZ.
In other words, convolution is adopted not only for efficiency but also because it is well matched to the underlying signal structure of collision-induced correlation features.
At the same time, compared to fully connected layers, 1D convolutional layers require significantly fewer parameters due to weight sharing, which helps keep the model lightweight despite the increased input size.
This is particularly attractive for practical NTN deployments where computational and memory resources may be limited.

The structure of our proposed early preamble collision classifier is shown in Fig.~\ref{fig:classifier}.
The input data consists of the correlation values of the preamble's ZCZ for each antenna, denoted as $c_r^j\left[m\right]$, where $m\in\left[\left(i-1\right)N_{\mathrm{CS}}, iN_{\mathrm{CS}}-1\right]$.
The input layer is of size $N_{\mathrm{ant}}\times N_{\mathrm{CS}}$, where $N_{\mathrm{CS}}$ is set to 8 according to the condition in~\eqref{eq:ZCZ_condition}~\cite{TVT23_TwoStage}.
The kernel size of the 1D convolutional layer is set to 3, which reflects the typical delay spread of the SatCom channel as reported in~\cite{38.811} and captures local dependencies across adjacent samples.
After the input layer, the data passes through a 1D convolutional layer with a kernel size of 3, stride 1, padding 1, and 16 channels.
The rectified linear unit (ReLU), defined as $\mathrm{ReLU}(x)=\max \left(0,x\right)$, is used as the activation function.
The next layer is another 1D convolutional layer with the same kernel size, stride, and padding, but with 32 channels, designed to extract high-level features of the preamble collision pattern.
After passing through the activation function, the data is fed into a fully connected layer for classification.
The output size of the fully connected layer is $K+1$, corresponding to the number of classes.
The final output values $z_k$ are processed using the softmax function $\sigma(z_k)\triangleq\frac{e^{z_k}}{\sum^K_{i=0}e^{z_i}}$ to compute the probability of each class.
Cross-entropy is used for the loss function, i.e.,
\begin{equation}
    \mathcal{L}=-\frac{1}{N_{\mathrm{B}}}\sum^{N_{\mathrm{B}}}_{n=1}\sum^K_{k=0}s_n\left(k\right)\log{\hat{s}_n\left(k\right)},
\end{equation}
where $N_\mathrm{B}$ represents the batch size and $s\left(k\right)$ and $\hat{s}\left(k\right)$ denote the one-hot encoded ground truth label and the output of the softmax function, $\sigma(z_k)$, respectively.
The network is trained using the Adam optimizer to minimize this loss.

In summary, the proposed classifier allows the SBS to detect and characterize preamble collisions immediately after Step 1, rather than waiting until later stages.
This early information opens the door to more efficient response strategies in subsequent steps. 
In what follows, we introduce a method that leverages this information to reduce unnecessary resource usage and improve access success probability.

\subsection{Opportunistic Transmission Scheme for Step 3}
Given the ability to detect collisions immediately after Step 1, it becomes possible to optimize how Step 3 transmissions are handled in the presence of such collisions.
This subsection introduces an opportunistic transmission scheme that aims to improve access success probability while preserving PUSCH resource efficiency.
While the early preamble collision classifier enables the SBS to detect preamble collisions at an early stage, it can only identify whether a collision has occurred—not which specific users used the collided preamble.
A straightforward baseline approach is for the SBS to withhold the RAR message for preambles identified as collided. 
In this case, users that do not receive the RAR message—those that used the preamble identified as collided by the SBS—follow a random backoff policy and reinitiate the random access procedure from Step 1.
This approach can reduce wasted use of PUSCH resources and shorten random access delay by eliminating unnecessary Step 3 transmissions that would otherwise result in collisions.
However, when all users involved in a detected collision are forced to back off, their current preamble transmissions are wasted, yielding no immediate access success.
Consequently, the improvement in random access success probability remains limited.
This motivates the need for a more efficient scheme that can recover access opportunities while maintaining resource efficiency.

\begin{figure*}[b]
\hrule
\vspace{4mm}
    \begin{equation}
        \begin{aligned}
        \mathbb{P}\left[\mathrm{Success}\mid\hat{k}\right] &\approx P\sum^K_{k=0}k\cdot \mathbb{P}\left[k\mid\hat{k}\right] - P^2\sum^K_{k=0} k\left(k-1\right)\cdot \mathbb{P}\left[k\mid\hat{k}\right] + \frac{P^3}{2}\sum^K_{k=0} k\left(k-1\right)\left(k-2\right)\cdot \mathbb{P}\left[k\mid\hat{k}\right] \\
        & = P\cdot \mathbb{E}\left[k\mid \hat{k}\right] - P^2\cdot \mathbb{E}\left[k\left(k-1\right)\mid \hat{k}\right] +\frac{P^3}{2}\cdot \mathbb{E}\left[k\left(k-1\right)\left(k-2\right)\mid \hat{k}\right]
        \end{aligned}
        \label{eq:Long}\tag{12}
    \end{equation}
\end{figure*}
\begin{figure*}[b]
\hrule
\vspace{4mm}
    \begin{equation}
        P^\star = \frac{2\mathbb{E}\left[k\left(k-1\right)\mid\hat{k}\right]\pm\sqrt{4\mathbb{E}\left[k\left(k-1\right)\mid\hat{k}\right]^2-6\mathbb{E}\left[k\left(k-1\right)\left(k-2\right)\mid\hat{k}\right]\cdot \mathbb{E}\left[k\mid\hat{k}\right]}}{3\mathbb{E}\left[k\left(k-1\right)\left(k-2\right)\mid\hat{k}\right]}, \quad 0\le P^\star \le 1
        \label{eq:OptimalP}\tag{14}
    \end{equation}
\end{figure*}

To address this limitation, we propose a proper Step 3 transmission scheme that leverages not only the preamble collision information but also the estimated number of collided users provided by the early preamble collision classifier. 
The core idea of the proposed scheme is to allow collided users to still have an opportunity to attempt Step 3 rather than forcing \textit{all} of them to back off.
To reflect the varying severity of collisions across different preambles, the SBS calculates a transmission probability for each preamble index based on the estimated number of collided users, where the probability is inversely related to the number of collisions.
This design ensures that, as the collision level increases, the chance of attempting Step 3 decreases, striking a balance between access opportunities and collision risk. 
Each user then decides whether to attempt Step 3 opportunistically according to the probability provided in the RAR message.
The detailed operation of the scheme is as follows:
\begin{enumerate}
    \item The SBS identifies the number of collided users in each ZCZ using the collision classifier.
    \item Based on the estimated number of collided users, the SBS computes the Step~3 transmission probability for each preamble index.
    \item The SBS transmits the RAR message to the users, including the preamble index, PUSCH grant, and a temporal identifier linking the RAR to the original preamble. 
    In addition, as a lightweight 3GPP standard extension, the SBS delivers an 8-bit quantized index representing the Step~3 transmission probability (e.g., via an explicit RAR extension field or a dedicated MAC control element scheduled together with the RAR), without altering the baseline random access timeline~\cite{38.321,38.331}.
    \item Each user refers to the RAR corresponding to its transmitted preamble index and decides whether to attempt Step~3 or perform random backoff based on the received probability index.
\end{enumerate}

Based on the estimated number of collided users, we now derive the Step 3 transmission probability that maximizes the random access success probability. 
Notably, the following derivation does not rely on any specific traffic arrival model. 
It is conditioned on the estimated collision multiplicity $\hat{k}$ for a given random access occasion, and thus remains valid regardless of how the set of active users is generated.
This probability allows the system to balance the trade-off between minimizing unnecessary Step 3 transmissions and maintaining a high success probability, even in the presence of collisions.
Let the number of collided users in the ZCZ of a given preamble be classified as $\hat{k}$, and let the Step 3 transmission probability for this preamble be denoted as $P$.
For notational simplicity, we omit the preamble index in the following derivation, without loss of generality. 
We define successful random access as the event in which exactly one of the users who selected the same preamble attempts Step 3, while all others perform backoff.
Based on this definition, the random access success probability for the preamble is given as
\begin{equation}
    \begin{aligned}
    \mathbb{P}\left[\mathrm{Success}\mid \hat{k}\right] = \sum^K_{k=0}k\cdot P\left(1-P\right)^{k-1}\cdot \mathbb{P}\left[k\mid\hat{k}\right],
    \end{aligned}\label{eq:MSG3_success}
\end{equation}
where $k$ represents the actual number of collided users.
The term $\left(1-P\right)^{k-1}$ can be approximated using the Taylor series expansion as
\begin{equation}
    \left(1-P\right)^{k-1} \approx 1-\left(k-1\right)P + \frac{\left(k-1\right)\left(k-2\right)}{2}P^2.
    \label{eq:Approximation}
\end{equation}
Substituting this approximation \eqref{eq:Approximation} into \eqref{eq:MSG3_success} yields the expression shown in~\eqref{eq:Long}.
Taking the derivative of \eqref{eq:Long} with respect to $P$ and setting it to zero yields the value of $P$ that maximizes the success probability:
\begin{equation}
    \begin{aligned}
    \frac{\mathrm{d}}{\mathrm{d}P}\mathbb{P}[\mathrm{Success}\mid \hat{k}]&=\mathbb{E}\left[k\mid\hat{k}\right]-2P\cdot \mathbb{E}\left[k\left(k-1\right)\mid\hat{k}\right]\\
    &\quad +\frac{3}{2}P^2\cdot \mathbb{E}\left[k\left(k-1\right)\left(k-2\right)\mid\hat{k}\right].
    \end{aligned}\tag{13}
\end{equation}
The final expression for the Step 3 transmission probability that maximizes the random access success probability is provided in \eqref{eq:OptimalP}.

To compute the optimal Step 3 transmission probability $P^\star$ in~\eqref{eq:OptimalP}, it is necessary to know the conditional probability $\mathbb{P}[k|\hat{k}]$.
Since the actual number of collided users $k$ is unknown, we approximate this term as:
\begin{align}
\mathbb{P}\left[k\mid\hat{k}\right] & = \frac{\mathbb{P}\left[\hat{k}\mid k\right]\cdot \mathbb{P}\left[k\right]}{\sum_{k=0}^{K}\mathbb{P}\left[\hat{k}\mid k\right]\cdot \mathbb{P}\left[k\right]} \label{eq:bayes_first} \tag{15}\\
& \approx \frac{Q\left(\hat{k},k\right)\cdot \mathbb{P}\left[k\right]}{\sum_{k=0}^{K}Q\left(\hat{k},k\right)\cdot \mathbb{P}\left[k\right]}. \label{eq:bayes_second} \tag{16}
\end{align}
Here, $Q(\hat{k},k)$ denotes the confusion matrix which represents the proportion of samples with true label class $k$ that are classified as class $\hat{k}$ during the training of the collision classifier. 
The remaining probability term $\mathbb{P}\left[k\right]$ can either be inferred from long-term statistics or more simply by assuming that all $D$ users randomly select a preamble from the $N_{\mathrm{PA}}$ available options and is expressed as follows:
\begin{equation}
    \mathbb{P}\left[k\right] \approx \binom{\hat{D}}{k} \left(\frac{1}{N_{\mathrm{PA}}}\right)^k\left(1-\frac{1}{N_{\mathrm{PA}}}\right)^{\hat{D}-k}.
    \label{eq:random_select} \tag{17}
\end{equation}
Here, $\hat{D}$ represents the estimated total number of active users, which is computed by summing the class indices output by the collision classifier across all preambles’ ZCZs.
Strictly speaking, the derived Step 3 transmission probability should be referred to as ``quasi-optimal'' because it depends on the classifier’s accuracy $Q(\hat{k},k)$ and the inherent uncertainty in estimating the total number of active users $\hat{D}$, rather than exact values.
By substituting \eqref{eq:random_select} into \eqref{eq:bayes_second}, we can obtain $\mathbb{P}[k|\hat{k}]$ which allows us to compute the expected values required for the quasi-optimal $P$ in \eqref{eq:OptimalP}. 
The overall operation of the proposed opportunistic transmission scheme is summarized in Algorithm~\ref{alg:opportunistic_msg3}.

The derivation of the Step 3 transmission probability $P^\star$ involves three approximations: (A1) second-order Taylor truncation of $(1-P)^{k-1}$ in \eqref{eq:Approximation}, (A2) confusion-matrix-based approximation of $P[k \mid \hat{k}]$ in \eqref{eq:bayes_second}, and (A3) substitution of $\hat{D}$ into the binomial prior $P[k]$ in \eqref{eq:random_select}.
The first approximation contributes negligible error within the operating regime, as the leading truncated term shifts $P^\star$ by less than $10^{-3}$.
The second approximation benefits from the ratio structure of $P^\star$ in \eqref{eq:OptimalP}, which makes the optimal probability first-order insensitive to multiplicative perturbations of the confusion matrix.
The third approximation is not a simplifying modeling choice but follows directly from the 3GPP-compliant uniform preamble selection~\cite{38.213}, with $|\hat{D} - D|$ remaining within a few users as supported by the classifier's reliable preamble presence detection (Fig.~\ref{fig:dataset_requirment}) and consistent per-class accuracy (Table~\ref{tab:classification_accuracy}).
The combined effect of these approximations is quantified empirically in Sec.\ V.B through a comparison with a genie-aided variant of the proposed framework.

\section{Simulation Results}\label{sec_5}
This section evaluates the performance of the proposed deep learning-based random access framework for LEO satellite communication systems.
The tapped delay line (TDL) channel models recommended for NTN link-level evaluation in TR 38.811~\cite{38.811} and TR 38.821~\cite{38.821} are employed throughout this section, consistent with the 3GPP-standardized evaluation methodology.
As mandated by 3GPP NR/NTN~\cite{38.811,38.821,38.213}, the user performs GNSS-based TA and Doppler pre-compensation prior to PRACH transmission, with residual offsets bounded by the user accuracy requirements in TS 38.133~\cite{38.133}.
Other random access-related parameters are summarized in Table~\ref{tab:sim_params}.

\begin{algorithm}[t]
\caption{Proposed Opportunistic Transmission Scheme}
\label{alg:opportunistic_msg3}
\textbf{Require:} Classifier outputs $\hat{k}$, confusion matrix $Q(\hat{k}, k)$, estimated number of active users $\hat{D}$

\begin{algorithmic}[1]
\Statex \textbf{SBS Procedure:}
\For{each preamble index}
    \State Compute $\mathbb{P}[k \mid \hat{k}]$ using \eqref{eq:bayes_second} and \eqref{eq:random_select}
    \State Derive $P$ by solving \eqref{eq:OptimalP}
    \State Include $P$ in the RAR message with the preamble index, PUSCH grant, and temporary identifier
\EndFor
\State Broadcast RAR messages to users

\Statex
\Statex \textbf{User Procedure:}
\For{each user that transmitted a preamble}
    \If{RAR message is received for its preamble}
        \State Attempt Step 3 transmission with probability $P$
    \Else
        \State Backoff and retry
    \EndIf
\EndFor
\end{algorithmic}
\end{algorithm}

\begin{figure}[t]
    \centering
    \includegraphics[width=1\columnwidth]{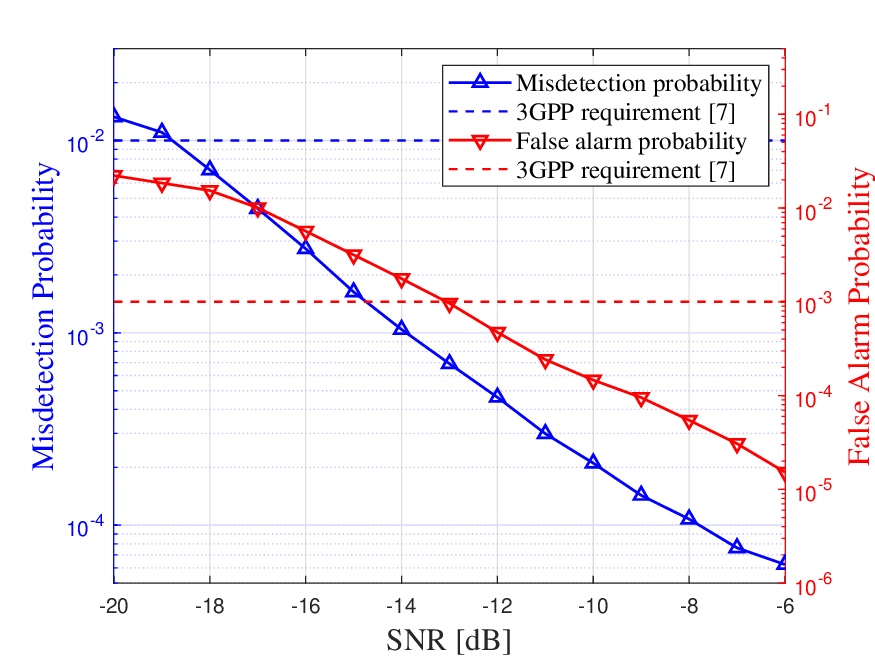}
    \caption{Misdetection and false alarm probabilities of the proposed preamble collision classifier, with models trained individually at each SNR.}
    \label{fig:dataset_requirment}
\end{figure}

The dataset for training the collision classifier consists of ZCZ correlation values generated under the TDL-B and TDL-D channel models, which are representative of non-LoS (NLoS) and LoS-dominant propagation environments, respectively.
We emphasize that the classifier is designed to learn collision-induced local patterns in the correlation sequence, particularly within the ZCZ, rather than to emulate every practical propagation impairment exhaustively.
Data are collected for all classes from 0 to $K$, corresponding to $K+1$ collision scenarios.
To accurately simulate the reception of preambles under these scenarios, appropriate transmit power and noise configurations are required.
In 3GPP-compliant systems, preamble transmission is subject to open-loop power control, where the transmit power is adjusted to achieve a target received power at the base station and thereby satisfy the detection criteria~\cite{38.213}.
To capture this in our simulation, we configure the transmit power and noise level based on 3GPP detection requirements.
Specifically, we consider the 3GPP-specified targets of misdetection probability below 1\% and false alarm probability below 0.1\% as a reference operating requirement~\cite{Book_RABO}.
\begin{table}[!t]
\centering
\caption{Simulation parameter settings~\cite{TVT23_TwoStage, TWC17_NORA, 36.213, IoTJ21_ERA, 38.821}.}
\label{tab:sim_params}
\begin{tabular}{ll}
\toprule
\textbf{Simulation Parameters} & \textbf{Value} \\
\midrule
Length of ZC sequence & $839$ \\
Bandwidth of PRACH [MHz] & $1.08$ \\
Size of cyclic shift & $8$ \\
Number of ZC root sequences & $1$ \\
Total number of available preambles & $104$ \\
Total number of RACH slots & $2000$ \\
Maximum RTT pre-compensation error [samples] & $2$ \\
RACH slot period [ms] & $10$ \\
Maximum number of random access retrial & $10$ \\
Transmission time for Step 1 [ms] & $1$ \\
Preamble detection processing time [ms] & $2$ \\
Transmission time for Step 2 [ms] & $1$ \\
Processing time between Step 2 and Step 3 [ms] & $3$ \\
Transmission time for Step 3 [ms] & $3$ \\
Transmission time for Step 4 [ms] & $1$ \\
Length of RAR window [ms] & $10$ \\
Length of CR window [ms] & $64$ \\
Length of backoff window [ms] & $20$ \\
\bottomrule
\end{tabular}
\end{table}
Fig.~\ref{fig:dataset_requirment} evaluates the misdetection and false alarm probabilities of the proposed classifier under various signal-to-noise ratio (SNR) conditions, where the misdetection probability is defined as the likelihood of classifying a non-zero class as class 0, and the false alarm probability is the likelihood of classifying a true class 0 as another class.
To satisfy 3GPP requirements~\cite{Book_RABO}, the minimum required SNR is approximately $-13$ dB.
Accordingly, 100,000 data samples were collected at each SNRs of \{$-13$, $-12$, $-11$, $-10$\} dB, of which 70\% were used for training and 30\% for testing.
Additionally, as in \cite{TWC21_DLRA}, to demonstrate the generalization performance of the trained classifier, 30,000 samples were collected at each SNRs of \{$-9$, $-8$, $-7$, $-6$, $-5$, $-4$\} dB as a separate test set.
Hyperparameters including learning rate, batch size, number of epochs, and Adam optimizer settings were tuned using $K$-fold cross-validation. 
The final values were set to a learning rate of $10^{-3}$, a batch size of 32, 20 epochs, and Adam optimizer parameters $(\beta_1, \beta_2) = (0.9, 0.999)$, respectively.

The following schemes were considered for comparison:
\begin{itemize}
    \item \textbf{Deep learning-based random access framework for terrestrial networks (DRA)~\cite{TWC21_DLRA}}:
    This scheme employs a deep learning-based classifier for collision detection and uses a separate neural network to extract TA information, which is then used to identify users and allocate distinct Step~3 grants.
    However, it assumes that TA is stationary and known to users in advance, making it unsuitable for LEO SatCom systems.
    Therefore, for a fair and implementable comparison under LEO SatCom conditions, we adopt an LEO-compatible adaptation of DRA as follows:
    when a collided preamble is detected, all users who transmitted that preamble are forced to back off (i.e., no Step~3 grant is allocated for that preamble in the current attempt).

    \item \textbf{Deep learning-based double contention random access (DCRA)~\cite{TWC23_DLDouble}}:
    This scheme extends DRA~\cite{TWC21_DLRA} by adopting a wider and deeper neural network for improved collision classification.
    When two users are classified as collided, the system allocates two Step~3 grants, and each user randomly selects one of them for transmission.

    \item \textbf{Single-root preamble sequence-based early collision detection (S-eCD)~\cite{TVT21_SeCD}}:
    This scheme uses a single root ZC sequence and assigns a distinct prime number and cyclic shift to each user such that their product maps to a unique correlation index.
    Since the BS can distinguish different prime numbers regardless of cyclic shift, users selecting the same prime number are identified as collided.
    The scheme detects collisions at an early stage and forces collided users to back off.

    \item \textbf{Multiple threshold-based early collision detection (MT-eCD)~\cite{TVT25_MTeCD}}:
    This scheme is designed for LoS-dominant LEO SatCom channels and detects early preamble collisions by exploiting the constructive and destructive interference behavior using multiple decision thresholds.
    The scheme detects collisions at an early stage and forces collided users to back off.
\end{itemize}

\subsection{Preamble Collision Classification Performance}
This subsection demonstrates the effectiveness of the proposed early preamble collision classifier in LEO satellite communication systems. 
For comparison, preamble collision classifiers originally developed for terrestrial networks, namely the DRA~\cite{TWC21_DLRA} and the DCRA~\cite{TWC23_DLDouble}, are adopted as benchmarks.

\begin{table*}[t]
\centering
\caption{Classification accuracy of preamble collision classifiers with varying channel models, numbers of antennas, and maximum number of classes ($K$).}
\label{tab:classification_accuracy}
\resizebox{\textwidth}{!}{
\begin{tabular}{lcccc|cccc|cccc|cccc}
\toprule
& \multicolumn{4}{c|}{1 antenna} & \multicolumn{4}{c|}{2 antennas} & \multicolumn{4}{c|}{4 antennas} & \multicolumn{4}{c}{8 antennas} \\
\cmidrule(lr){2-5} \cmidrule(lr){6-9} \cmidrule(lr){10-13} \cmidrule(lr){14-17}
& $K=3$ & $K=4$ & $K=5$ & $K=6$ & $K=3$ & $K=4$ & $K=5$ & $K=6$ & $K=3$ & $K=4$ & $K=5$ & $K=6$ & $K=3$ & $K=4$ & $K=5$ & $K=6$ \\
\midrule
\multicolumn{17}{l}{\textbf{(a) TDL-B}} \\
DRA~\cite{TWC21_DLRA} & 0.774 & 0.678 & 0.598 & 0.536 & 0.825 & 0.727 & 0.649 & 0.591 & 0.862 & 0.777 & 0.700 & 0.644 & 0.898 & 0.830 & 0.767 & 0.714 \\
DCRA~\cite{TWC23_DLDouble} & 0.775 & 0.680 & 0.604 & 0.544 & 0.833 & 0.740 & 0.659 & 0.599 & 0.865 & 0.784 & 0.703 & 0.648 & 0.902 & 0.837 & 0.771 & 0.716 \\
\textbf{Proposed} & \textbf{0.773} & \textbf{0.675} & \textbf{0.591} & \textbf{0.541} & \textbf{0.830} & \textbf{0.739} & \textbf{0.658} & \textbf{0.597} & \textbf{0.863} & \textbf{0.778} & \textbf{0.700} & \textbf{0.642} & \textbf{0.900} & \textbf{0.831} & \textbf{0.764} & \textbf{0.710} \\
\midrule
\multicolumn{17}{l}{\textbf{(b) TDL-D}} \\
DRA~\cite{TWC21_DLRA}  & 0.849 & 0.754 & 0.669 & 0.608 & 0.883 & 0.807 & 0.731 & 0.660 & 0.917 & 0.854 & 0.791 & 0.732 & 0.951 & 0.912 & 0.864 & 0.813 \\
DCRA~\cite{TWC23_DLDouble} & 0.850 & 0.755 & 0.678 & 0.609 & 0.884 & 0.804 & 0.736 & 0.668 & 0.915 & 0.855 & 0.793 & 0.732 & 0.952 & 0.915 & 0.865 & 0.822 \\
\textbf{Proposed} & \textbf{0.845} & \textbf{0.751} & \textbf{0.667} & \textbf{0.603} & \textbf{0.897} & \textbf{0.824} & \textbf{0.746} & \textbf{0.681} & \textbf{0.941} & \textbf{0.884} & \textbf{0.820} & \textbf{0.759} & \textbf{0.972} & \textbf{0.930} & \textbf{0.882} & \textbf{0.835} \\
\bottomrule
\end{tabular}
}
\end{table*}

\begin{table*}[t]
\centering
\caption{Number of parameters for each classifier with varying numbers of antennas and maximum number of classes ($K$).}
\label{tab:num_parameters}
\resizebox{\textwidth}{!}{
\begin{tabular}{lcccc|cccc|cccc|cccc}
\toprule
& \multicolumn{4}{c|}{1 antenna} & \multicolumn{4}{c|}{2 antennas} & \multicolumn{4}{c|}{4 antennas} & \multicolumn{4}{c}{8 antennas} \\
\cmidrule(lr){2-5} \cmidrule(lr){6-9} \cmidrule(lr){10-13} \cmidrule(lr){14-17}
& $K=3$ & $K=4$ & $K=5$ & $K=6$ & $K=3$ & $K=4$ & $K=5$ & $K=6$ & $K=3$ & $K=4$ & $K=5$ & $K=6$ & $K=3$ & $K=4$ & $K=5$ & $K=6$ \\
\midrule
DRA~\cite{TWC21_DLRA}      & 1.766e5 & 1.769e5 & 1.771e5 & 1.774e5 & 1.766e5 & 1.769e5 & 1.771e5 & 1.774e5 & 1.766e5 & 1.769e5 & 1.771e5 & 1.774e5 & 1.766e5 & 1.769e5 & 1.771e5 & 1.774e5 \\
DCRA~\cite{TWC23_DLDouble} & 6.270e6 & 6.271e6 & 6.272e6 & 6.273e6 & 6.270e6 & 6.271e6 & 6.272e6 & 6.273e6 & 6.270e6 & 6.271e6 & 6.272e6 & 6.273e6 & 6.270e6 & 6.271e6 & 6.272e6 & 6.273e6 \\
\textbf{Proposed}                   & \textbf{1,844}   & \textbf{1,909}   & \textbf{1,974}   & \textbf{2,039}   & \textbf{1,940}   & \textbf{2,005}   & \textbf{2,070}   & \textbf{2,135}   & \textbf{2,036}   & \textbf{2,101}   & \textbf{2,166}   & \textbf{2,231}   & \textbf{2,228}   & \textbf{2,293}   & \textbf{2,358}   & \textbf{2,423} \\
\bottomrule
\end{tabular}
}
\end{table*}

Table~\ref{tab:classification_accuracy} presents the classification accuracy of the preamble collision classifiers under different channel models, antenna configurations, and values of $K$.
Recall that the classifier has $(K+1)$ output classes, corresponding to 0 through $K$ users selecting the same preamble.
Under the TDL-B channel model, all three classifiers exhibit similar classification accuracy.
Since TDL-B is a NLoS channel model, it tends to yield less pronounced local structures in the ZCZ region, which can reduce the benefit of the 1D convolutional layer.
As reflected in Table~\ref{tab:classification_accuracy}, in such cases, deeper and wider neural networks, as used in DRA and DCRA, may slightly improve classification by compensating for the weak structural features.
In contrast, under the TDL-D channel model, which is dominated by LoS propagation, a different trend emerges, which is representative of LoS-dominant settings often considered in LEO SatCom scenarios.
The proposed classifier consistently outperforms the others in terms of classification accuracy.
This improvement can be attributed to the more prominent peak observed in the ZCZ region under LoS-dominant channels, which makes the correlation value patterns around the peak more distinguishable when a collision occurs.
Moreover, the performance gap widens as the number of antennas increases.
This is because DRA and DCRA average correlation values across antennas to enhance stability, whereas the proposed classifier uses all per-antenna correlation values, allowing it to exploit richer spatial information.

Table~\ref{tab:num_parameters} presents the number of parameters required by each preamble collision classifier for different numbers of antennas and class settings ($K$).
For DRA and DCRA, the number of parameters remains constant regardless of the number of antennas, as the input data size does not vary with antenna count.
On the other hand, the proposed classifier’s parameter count increases with the number of antennas due to the expanded input size.
However, while DRA and DCRA use only fully connected layers with dense inter-node connections, the proposed classifier primarily employs convolutional filters.
This design drastically reduces the number of required parameters.
For instance, with 8 antennas and $K=6$, the proposed classifier requires about 73 times fewer parameters than DRA and nearly 2,600 times fewer parameters than DCRA.
This reduction directly translates into lower on-board compute and memory requirements, which is particularly beneficial for practical NTN deployments subject to stringent payload constraints~\cite{WCM24_LEOhardware}.
In summary, the proposed classifier provides a favorable accuracy--complexity trade-off, achieving comparable or even superior classification accuracy with substantially reduced model size and implementation cost.

Beyond parameter count, we further quantify the actual on-board computational feasibility of the proposed framework.
Counting multiplications and additions, the inference cost of the proposed classifier is approximately $10.4$ kFLOPs per preamble.
For $N_{\mathrm{PA}} = 104$ available preambles, the total inference cost per random access occasion is approximately $1.08$ MFLOPs.
This computation must complete within the RAR window of 10\,ms, translating to a required processing throughput of approximately $108$ MFLOPS.
As a representative on-board reference, a radiation-hardened FPGA such as the Xilinx Kintex UltraScale XQRKU060~\cite{Xilinx_XQRKU060} provides processing throughput on the order of TFLOPS for fixed-point inference, leaving an operational margin of approximately four orders of magnitude relative to the proposed framework's requirement.
The subsequent $P^\star$ computation in~\eqref{eq:OptimalP} involves only a per-preamble quadratic-equation evaluation, whose cost is negligible compared with the classifier inference.
The corresponding computational requirements of DRA and DCRA are summarized in Table~\ref{tab:flops_comparison} for comparison.
Therefore, the per-preamble computational overhead of the proposed framework is comfortably accommodated within the standard RAR timing budget on representative on-board hardware.

\subsection{Random Access Performance}
This subsection presents the end-to-end random access performance of the proposed framework.
To ensure a fair comparison across different schemes, all approaches were configured with the same number of available preambles, thereby maintaining an equal number of scheduled PUSCH resources.
However, in the case of DCRA, additional PUSCH resources are allocated to certain users when a preamble collision is detected. 
More specifically, if two users select the same preamble, DCRA assigns two separate PUSCH resources to that preamble to resolve the collision. 
As a result, even with the same number of available preambles, the total number of scheduled PUSCH resources in DCRA differs, since multiple grants may be allocated for a single preamble.
For the S-eCD scheme, both the ideal case with zero RTT uncertainty and the practical case with the same RTT uncertainty setting as other methods were evaluated.
Unless otherwise specified, all results are obtained under the configuration of 8 receive antennas, the TDL-D channel model and an SNR of 10\,dB.
Following the specification in~\cite{TWC23_DLDouble}, the DCRA scheme uses $K = 3$, whereas the proposed framework adopts $K = 6$.
The number of available preambles is given by $N_{\mathrm{PA}} = N_{\mathrm{R}} \lfloor N_{\mathrm{ZC}}/N_{\mathrm{CS}}\rfloor$ under the NTN configuration~\cite{38.211,TVT23_TwoStage,TVT25_MTeCD}.
The number of active users is set to a maximum of 52 by accounting for the LEO SatCom beam footprint size and the users’ access-attempt interval, which together determine the plausible number of simultaneous access attempts within a beam~\cite{38.821,36.763,REF_5MIN_IOT_REPORTING}.

\begin{table}[t]
\centering
\caption{Computational complexity comparison for each classifier.}
\label{tab:flops_comparison}
\resizebox{\columnwidth}{!}{
\begin{tabular}{lccc}
\toprule
& \textbf{DRA~\cite{TWC21_DLRA}} & \textbf{DCRA~\cite{TWC23_DLDouble}} & \textbf{Proposed} \\
\midrule
Per-preamble cost & $\sim$355 kFLOPs & $\sim$12.5 MFLOPs & $\sim$10.4 kFLOPs \\
Per-occasion cost & $\sim$37 MFLOPs & $\sim$1.3 GFLOPs & $\sim$1.08 MFLOPs \\
Required throughput & $\sim$3.7 GFLOPS & $\sim$130 GFLOPS & $\sim$108 MFLOPS \\
Operational margin~\cite{Xilinx_XQRKU060} & $\sim$$270\times$ & $\sim$$8\times$ & $\sim$$10{,}000\times$ \\
\bottomrule
\end{tabular}
}
\end{table}

\begin{figure}[t]
    \centering
    \includegraphics[width=1\columnwidth]{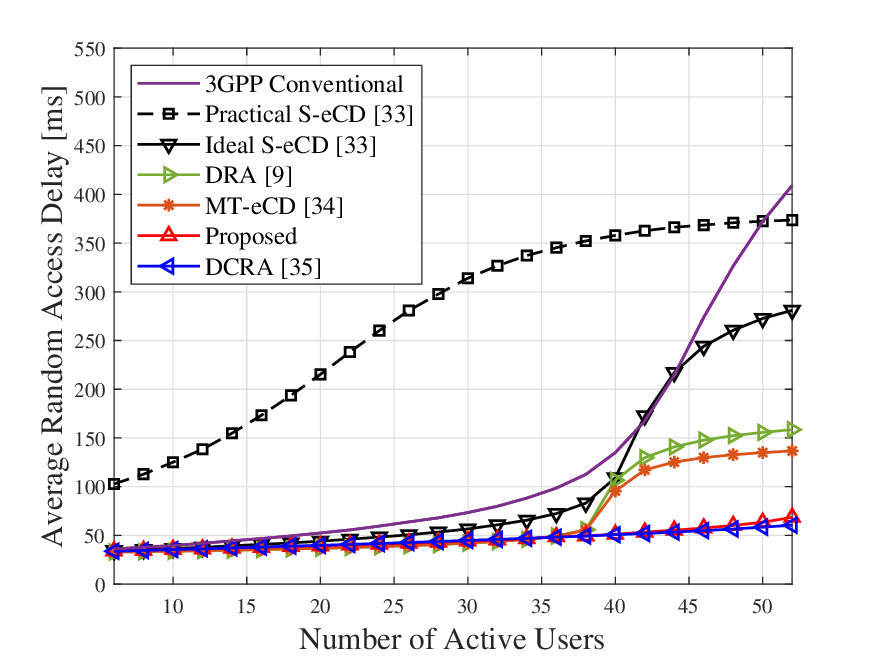}
    \caption{Average random access delay versus the number of active users. Random access delay is defined as the time elapsed from a user's initial access attempt to a successful random access. The total number of available preambles is 104.}
    \label{fig:delay}
\end{figure}

Fig.~\ref{fig:delay} illustrates the average random access delay for each scheme, defined as the time elapsed until a user successfully completes the random access procedure or exhausts the maximum number of retrials. 
This delay is averaged over mixed realizations that naturally include both collision-free and collision-present cases under the considered traffic and channel settings.
The practical S-eCD yields the longest delay due to limited early collision detection capability, whereas the ideal S-eCD achieves shorter delay through earlier backoff decisions. 

DRA and MT-eCD exhibit short delays at low user densities because they enforce immediate backoff upon collision detection, but the delay increases rapidly under heavy contention due to repeated deferrals. 
Compared with DCRA, the proposed framework achieves lower delay in low-density regimes by allowing collided users to probabilistically defer Step~3 after Step~2, while DCRA mandates Step~3 transmissions with additional PUSCH grants. 
As the user density increases, higher-order collisions become more frequent, and the two schemes exhibit different trade-offs between preserving access opportunities for collided users and suppressing repeated transmissions, which can lead to a crossing point in delay performance.
For reference, we also include the conventional 3GPP random access baseline, confirming that early preamble collision detection does not incur a delay penalty in the considered operating range.

\begin{figure}[t]
    \centering
    \includegraphics[width=1\columnwidth]{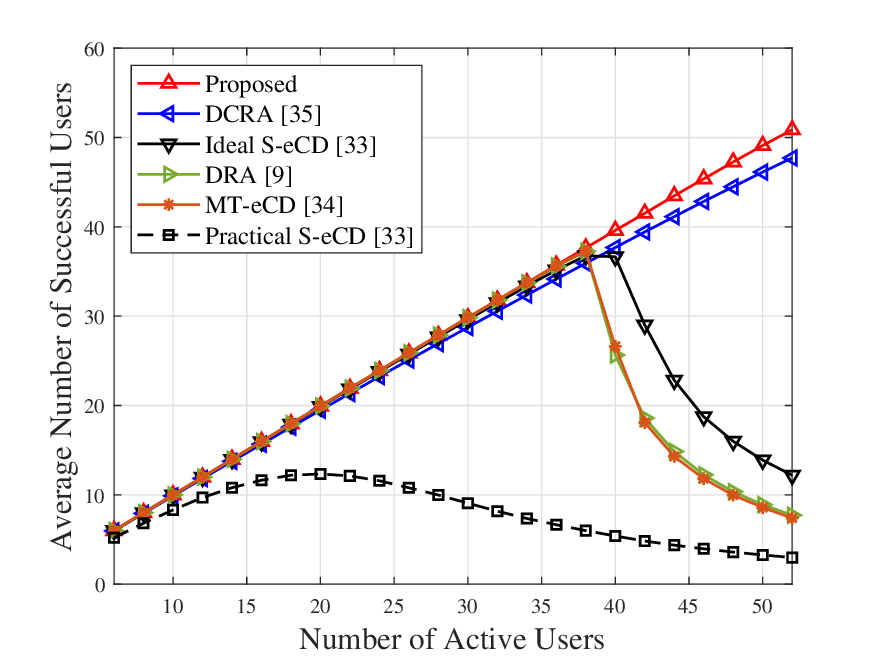}
    \caption{Average number of successful random access users versus the number of active users. Successful random access is defined as the case where the BS decodes Step 3 signals without collision. The total number of available preambles is 104.}
    \vspace{-4mm}
    \label{fig:success_nodes}
\end{figure}

Fig.~\ref{fig:success_nodes} shows the average number of successful random access users. 
In general, when the number of successful users increases nearly linearly with the number of active users, it implies that most users successfully complete the random access procedure, which is desirable. 
From this perspective, schemes that rely on immediate backoff upon collision detection, without additional contention-resolution actions, tend to suffer a sharp performance degradation as the user density increases. 
In contrast, both the proposed framework and DCRA sustain higher success performance by leveraging early collision awareness and resolving contention beyond simple backoff decisions. 
In particular, the proposed framework tends to achieve a higher number of successful accesses in the low-to-high density regime, because its collision-aware opportunistic Step~3 transmission provides additional access opportunities for collided users without deterministically discarding all of them.

\begin{figure}[t!]
    \centering
    \includegraphics[width=1\columnwidth]{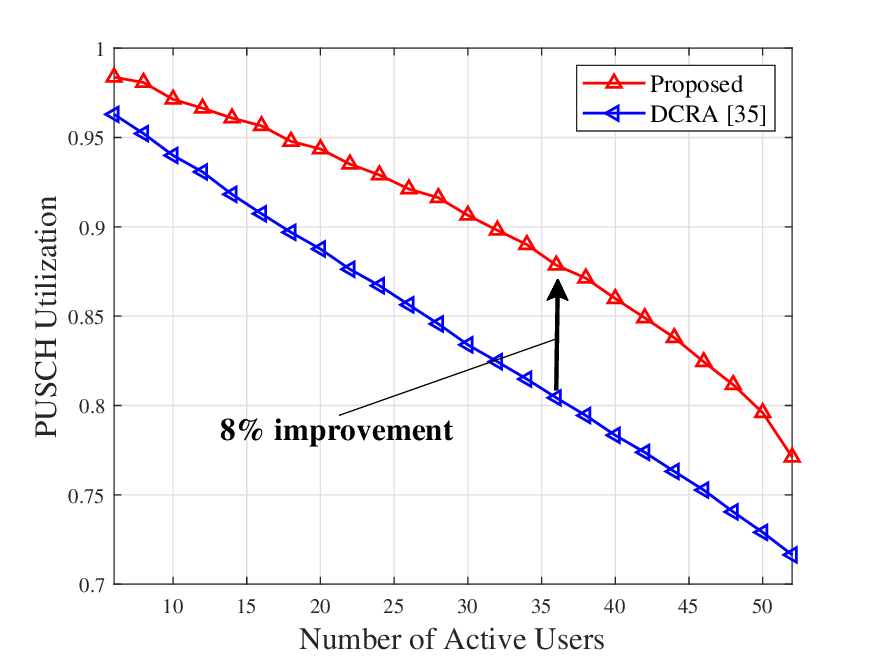}
    \caption{PUSCH utilization versus the number of active users. PUSCH utilization is defined as the fraction of Step 3 PUSCH resources successfully used without collision. The total number of available preambles is 104.}
    \label{fig:pusch_efficiency}
\end{figure}

The superiority of the proposed framework over DCRA is also evident in terms of PUSCH utilization. 
PUSCH utilization is defined as the ratio of the Step~3 PUSCH resources that are successfully used without collision to the total PUSCH resources allocated for Step~3. 
As shown in Fig.~\ref{fig:pusch_efficiency}, the proposed framework achieves higher PUSCH utilization than DCRA. 
This is because, in DCRA, when two users select the same preamble in Step~1, the base station allocates additional Step~3 PUSCH resources, resulting in two PUSCH resources being associated with a single preamble and enabling contention-based transmissions in Step~3. 
If the two users transmit on different PUSCH resources, no PUSCH resource is wasted; however, if both users attempt Step~3 on the same PUSCH resource, then the two allocated PUSCH resources are effectively wasted due to the collision. 
In contrast, the proposed framework does not allocate additional Step~3 PUSCH resources. 
Instead, users attempt Step~3 opportunistically based on the collision information, which reduces unnecessary PUSCH allocations and yields higher resource efficiency.

\begin{figure}[!t]
    \centering
    \includegraphics[width=1\columnwidth]{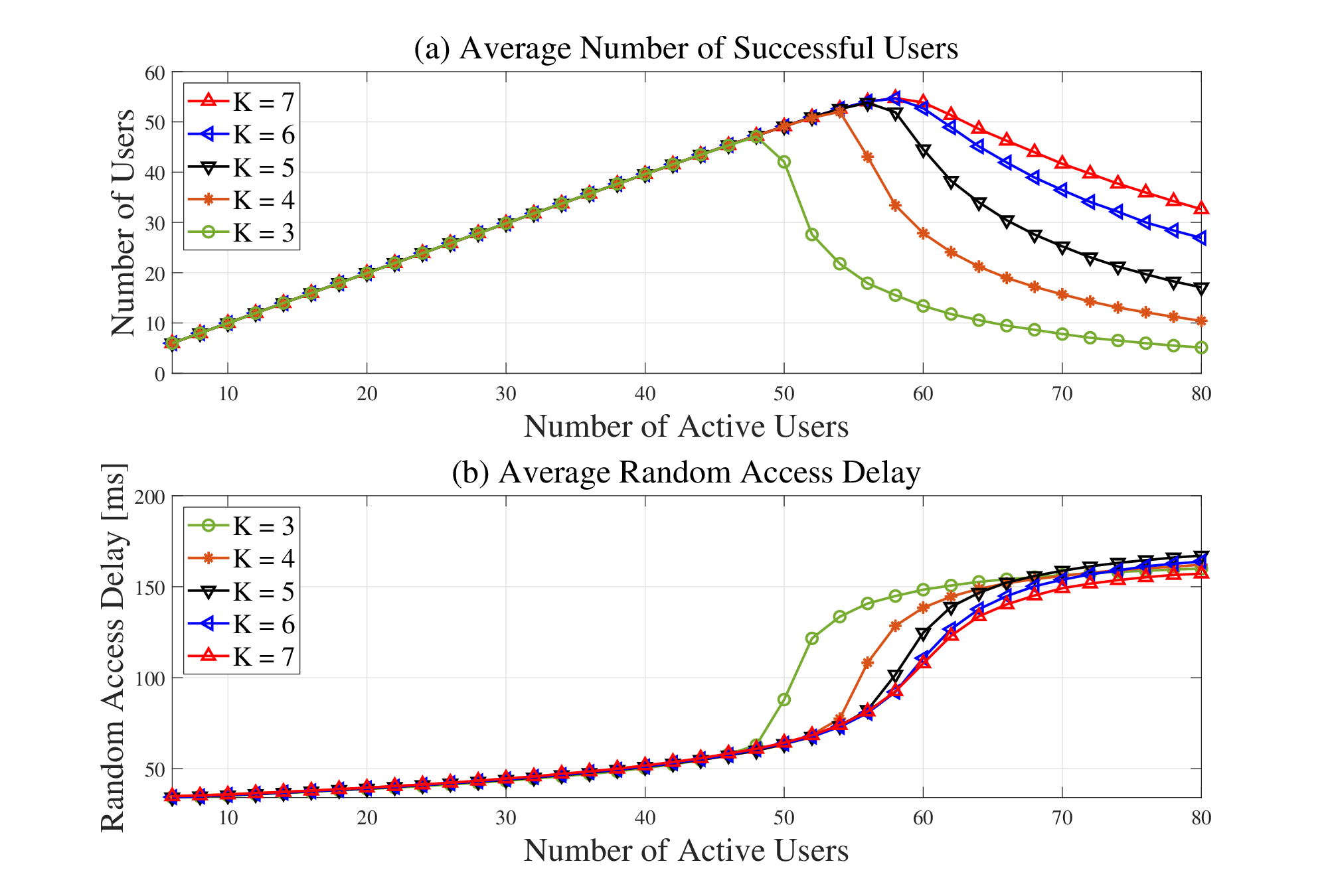}
    \caption{Performance of the proposed framework versus the number of active users under different $K$: (a) average number of successful users and (b) average random access delay.}
    \label{fig:class}
\end{figure}

Fig.~\ref{fig:class} investigates the impact of $K$ (the maximum collision multiplicity considered by the classifier) on the proposed framework. 
As $K$ increases, the framework can differentiate a wider range of collision multiplicities and apply more fine-grained Step~3 attempt control, which increases the number of successful users and delays the onset of performance saturation as the user density grows. 
Consistently, the random access delay is reduced over a wider user-density range because more collision cases can be handled without immediately forcing retransmission/backoff. 
However, the performance gain from increasing $K$ eventually saturates, indicating diminishing returns beyond a certain $K$ while the classification task becomes more complex due to the increased number of collision scenarios.

\begin{figure}[t!]
    \centering
    \includegraphics[width=1\columnwidth]{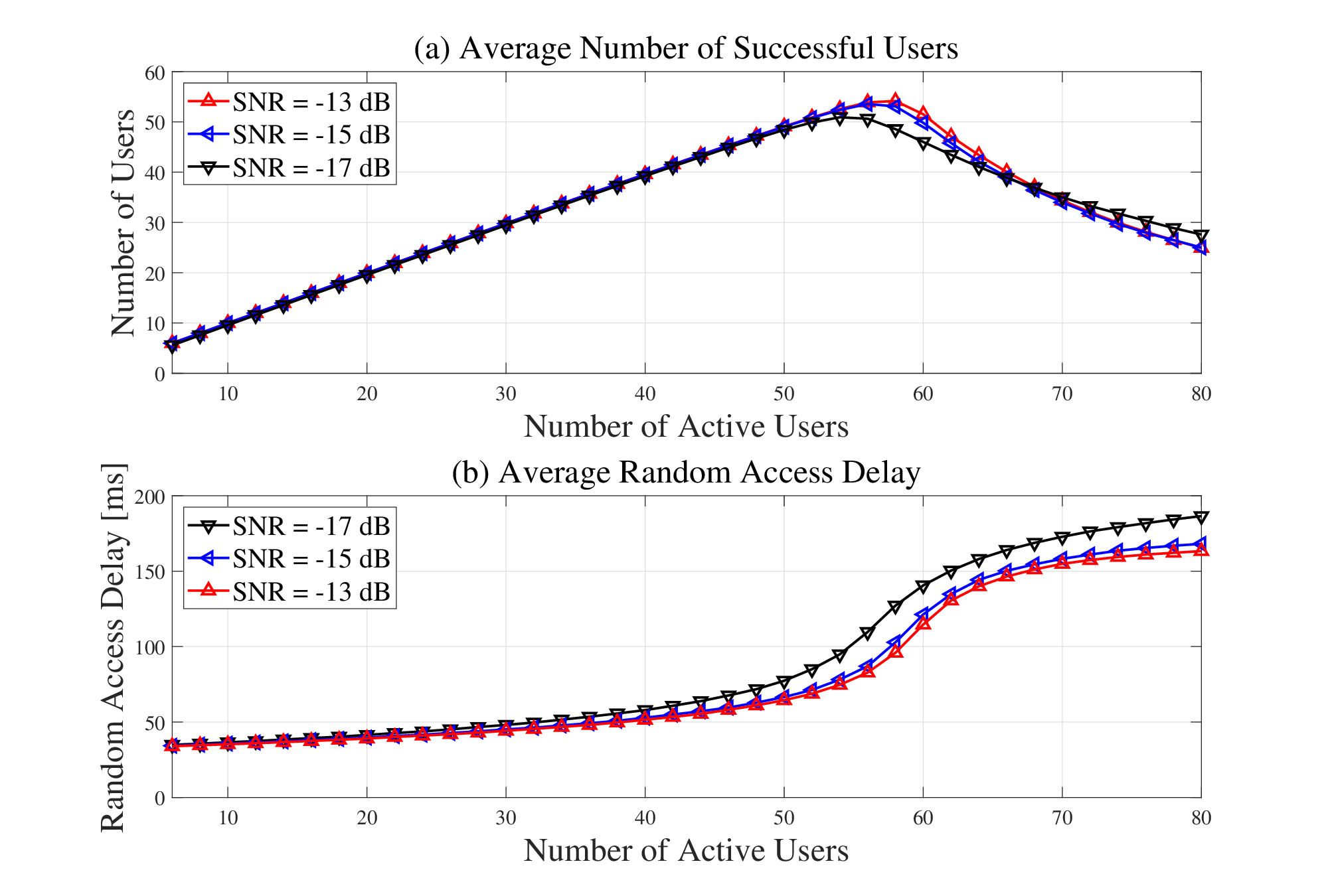}
    \caption{Performance of the proposed framework versus the number of active users under different SNRs: (a) average number of successful users and (b) average random access delay.}
    \label{fig:low_snr}
\end{figure}

Fig.~\ref{fig:low_snr} evaluates the proposed framework in a lower-SNR regime, $\text{SNR}\in[-17,-15,-13]$ dB, to assess robustness.
As the SNR decreases, the average number of successful users decreases and the average random access delay increases, since collision signatures in the correlation sequence become less distinguishable and misclassification becomes more likely.
Nevertheless, the proposed framework shows consistent trends with respect to user density across SNRs, indicating robust behavior rather than unstable performance.
These results also serve as a proxy for large-scale attenuation effects (e.g., rain attenuation and shadowing), which mainly shift the effective operating SNR without fundamentally changing the collision-induced correlation structures.

\begin{figure}[t!]
    \centering
    \includegraphics[width=1\columnwidth]{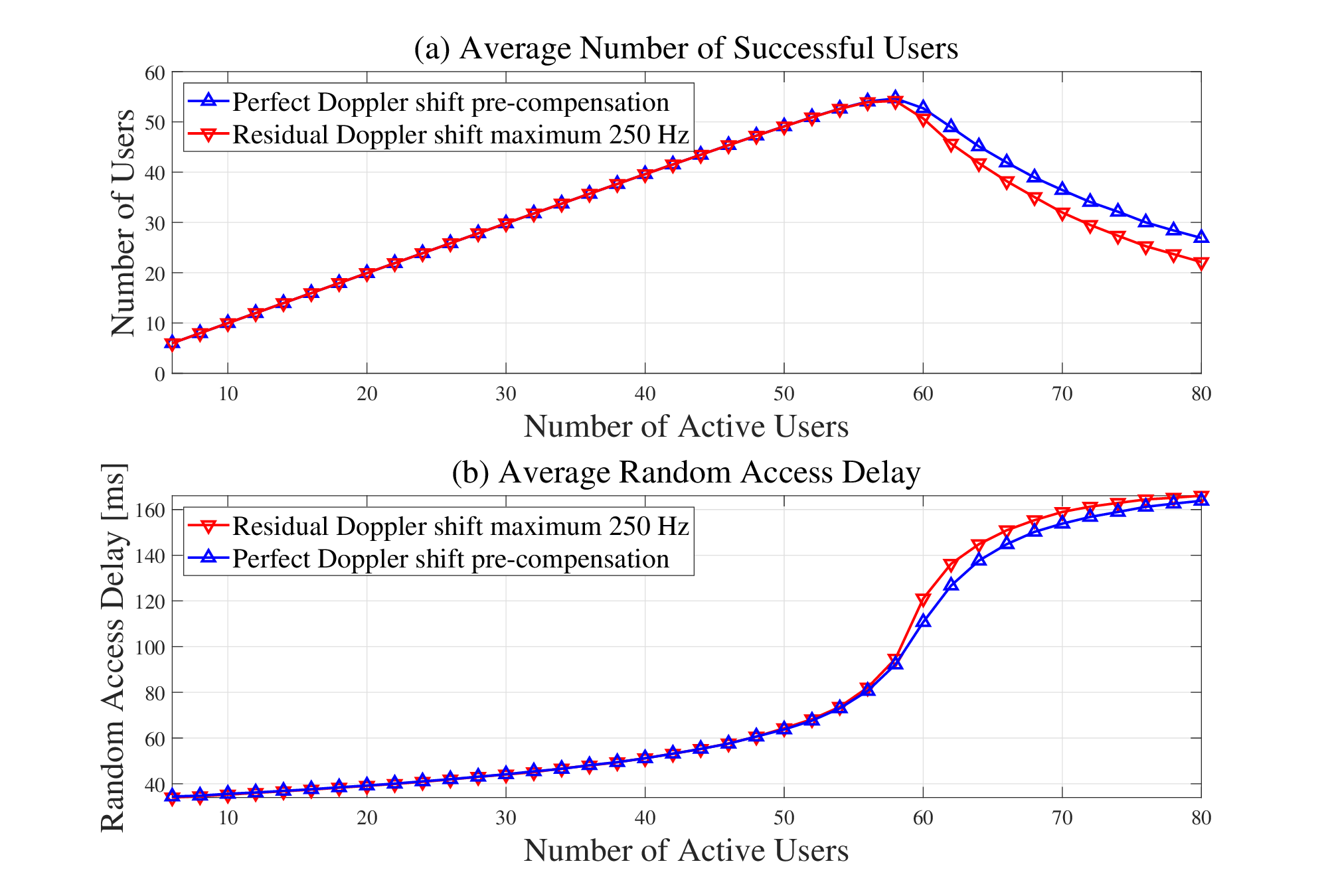}
    \caption{Performance of the proposed framework versus the number of active users under residual Doppler: (a) average number of successful users and (b) average random access delay.}
    \label{fig:doppler}
\end{figure}

Fig.~\ref{fig:doppler} examines the sensitivity of the proposed framework to residual Doppler (residual frequency offset) after Doppler compensation/tracking. 
According to 3GPP assumptions for frequency error after compensation, the residual offset can be on the order of 10~Hz~\cite{3GPP_FreqEst} however, to stress-test the proposed classifier and avoid optimistic conclusions, we additionally consider a much harsher case with a maximum residual Doppler of 250~Hz. 
Increased residual Doppler introduces phase rotation across the correlation processing interval, which blurs the collision-dependent structures and reduces the reliability of collision-multiplicity estimation. 
As a result, fewer users complete random access successfully and the access delay increases due to additional contention-resolution and retransmission events. 
Despite this stress-test setting, the degradation remains limited and the user-density-dependent trends are preserved, supporting the robustness of the proposed framework against residual Doppler.

To quantify the combined effect of the approximations introduced in the derivation of $P^\star$ (Sec.\ IV.C), we additionally evaluate a genie-aided variant of the proposed framework, in which $P^\star$ is computed from the true collision multiplicity $k$ and the true number of active users $D$, thereby removing the confusion-matrix-based posterior approximation and the $\hat{D}$-based binomial approximation.
Rev-Fig.\ 13(a) compares the average number of successful random access users, and Rev-Fig.\ 13(b) compares the average random access delay, between the practical framework and the genie-aided variant.
The gap in the average number of successful users is negligible in the low-to-moderate user-density range and remains within approximately 3 users (corresponding to about 10\% relative degradation) even in the high-density regime above 60 active users.
A similar pattern is observed for the average random access delay, where the gap is negligible up to about 60 active users and remains within approximately 15\,ms in the high-density regime.
We note that the high-density regime above 60 users lies beyond the saturation point of the proposed framework at the configured $K=6$, so the gap there reflects the behavior near the framework's intrinsic capacity limit rather than the operating regime targeted by the proposed design.
These results indicate that the practical framework operates close to the oracle upper bound within its targeted operating regime, confirming that the combined impact of the approximations introduced in Sec.\ IV.C is marginal at the system level.

Overall, these results demonstrate that the proposed framework effectively improves end-to-end random access performance by leveraging early collision awareness and collision-multiplicity-guided Step~3 control.
Across a wide range of user densities, it provides a favorable success--delay trade-off compared with the considered baselines, while maintaining robust performance trends under different design and impairment conditions, including the choice of $K$, low-SNR regimes, residual Doppler, and the approximations introduced in the derivation of $P^\star$.

\begin{figure}[!t]
    \centering
    \includegraphics[width=1\columnwidth]{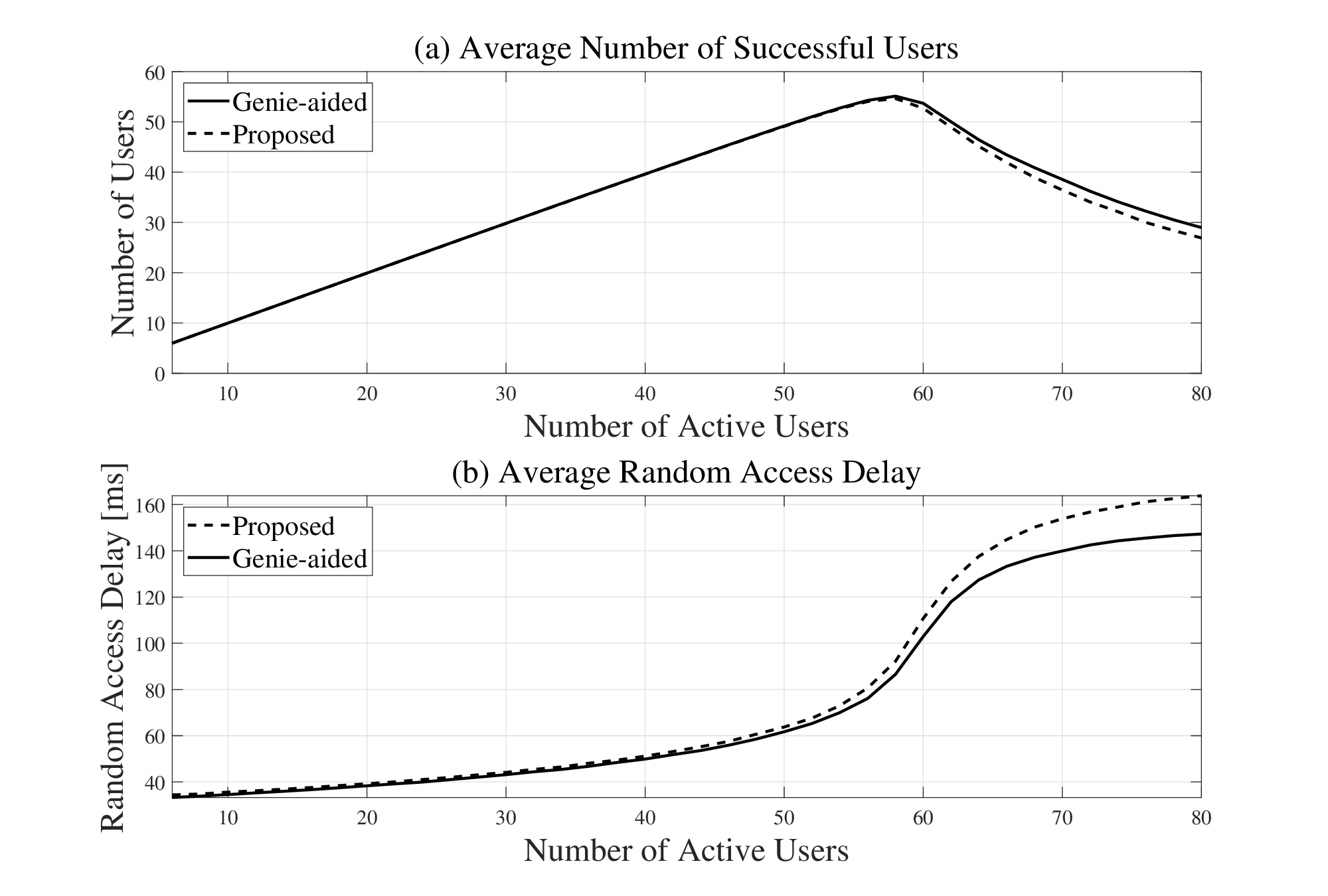}
    \caption{Performance of the proposed framework versus its genie-aided variant in terms of (a) average number of successful users and (b) average random access delay.}
    \label{fig:Genie}
\end{figure}

\subsection{Design Guidelines for NTN Deployment}

The evaluation results in Sec.\ V.B yield concrete design rules for configuring the proposed framework under given deployment expectations.
We extract three guidelines covering the choice of the classifier parameter $K$, the operating SNR range, and the residual Doppler tolerance.

\vspace{0.3em}
\noindent\textbf{Guideline 1 (Selection of $K$).}
The maximum collision multiplicity $K$ controls how many distinct collision cases the classifier can resolve, and thus directly affects the active-user load at which the random access performance begins to saturate.
From the saturation behavior in Fig.\ 10, the proposed framework sustains successful operation up to approximately $49$, $55$, $57$, $58$, and $59$ active users for $K = 3, 4, 5, 6, 7$, respectively.
The marginal gain in supported load diminishes rapidly beyond $K = 5$, while the classification task becomes more complex and the parameter count grows.
A practical design rule is therefore to choose the smallest $K$ such that the saturation point exceeds the expected peak active-user load with a small safety margin.
For the operating range considered in this work (peak load $\leq 50$), $K = 6$ provides full coverage with negligible marginal cost, while values of $K \geq 7$ yield diminishing marginal returns.

\vspace{0.3em}
\noindent\textbf{Guideline 2 (Operating SNR range).}
Fig.\ 11 evaluates the framework under SNRs from $-17$ to $-13$\,dB.
The framework maintains consistent user-density-dependent trends across this range, but the absolute number of successful users gradually decreases as the operating SNR drops below the 3GPP-recommended regime of $-13$\,dB~\cite{38.211}.
Accordingly, the recommended deployment configuration is to leverage the standard 3GPP open-loop power control~\cite{38.213} so that the effective operating SNR remains at or above $-13$\,dB, where the classifier satisfies the 3GPP misdetection and false-alarm requirements (Fig.\ 6).
For coverage-edge or heavily attenuated conditions where the effective SNR may fall below this regime, additional coverage-enhancement features such as message-1 repetition supported in 3GPP Rel-18~\cite{38.321} can be applied to restore the operating SNR within the recommended range.

\vspace{0.3em}
\noindent\textbf{Guideline 3 (Residual Doppler tolerance).}
Fig.\ 12 shows that the framework maintains stable performance under residual Doppler up to 250\,Hz, which is approximately $25\times$ larger than the typical post-compensation residual of $10$\,Hz reported in~\cite{3GPP_FreqEst}.
This indicates that the framework provides a substantial operational margin under residual frequency offset, and that conservative Doppler-tracking configurations at the user side are sufficient for stable operation.
Designers therefore do not need to push Doppler estimation accuracy beyond what is required by 3GPP NTN~\cite{38.811}; the framework remains robust under significantly larger residual Doppler than the typical post-compensation level.

\section{Conclusion}\label{sec_6}
This paper proposed a deep learning-based random access framework tailored for LEO SatCom systems. 
The framework incorporates an early preamble collision classifier that exploits antenna-wise correlation features to estimate the number of collided users at the initial stage of random access. 
A lightweight 1D convolutional neural network architecture enables efficient extraction of spatial and temporal features, even under constrained ZCZ. 
Based on the classifier’s output, an opportunistic transmission scheme was designed, where each user probabilistically attempts Step 3 transmission according to a quasi-optimal access probability. 
This probability reflects the estimated severity of collisions and balances the trade-off between collision risk and resource utilization. 
Simulation results validated that the proposed framework significantly improves access success probability, reduces delay, and increases PUSCH utilization under 3GPP-compliant LEO settings. 
Compared to conventional schemes, the proposed scheme achieves better performance with lower computational complexity, making it suitable for practical LEO SatCom deployments. 

Future research may extend this work by incorporating inter-satellite interference under more aggressive random access resource reuse across satellites/beams. 
Another promising direction is to integrate PHY-layer collision-type estimation with learning-based (e.g., RL) access-control policies for cross-layer contention management under dynamic traffic and coverage conditions.

\bibliographystyle{IEEEtran}
\bibliography{reference}

%\newpage

%\section*{Biography}

%\begin{IEEEbiographynophoto}{Hyunwoo Lee} (Graduate Student Member, IEEE) received a B.S. degree in electrical and electronic engineering from Yonsei University, Seoul, South Korea, in 2021, where he is currently pursuing his Ph.D. degree in electrical and electronic engineering. His research interests include MIMO systems, channel estimation, and LEO SatCom systems.

%\begin{IEEEbiographynophoto}{Daesik Hong} (Fellow, IEEE)  is a Professor at the School of Electrical and Electronic Engineering, Yonsei University. He has been serving as Chair of Samsung-Yonsei Research Center for Mobile Intelligent Terminals. He served as Vice President of Research Affairs and President of the Industry-Academic Cooperation Foundation, Yonsei University, from 2010 to 2011. He served as Chief Executive Officer (CEO) for Yonsei Technology Holding Company in 2011 and served as President of the Institute of Electronics and Information Engineers (IEIE) in 2017. From 2016 to 2020, he served as Dean of the College of Engineering, Yonsei University. He served as an editor of the IEEE Transactions on Wireless Communications and IEEE Wireless Communications Letters from 2006 to 2011 and from 2011 to 2016, respectively. His current research activities are focused on future wireless communication including 5G and 6G systems, OFDM(A), NOMA, in-band full-duplex, cognitive-radio, V2X, and satellite communications.

\end{document}